\documentclass[manuscript]{acmart}

\usepackage[ruled,lined]{algorithm2e}
\usepackage{xspace}
\usepackage{subfig}
\usepackage[dvipsnames]{xcolor}
\usepackage{tikz}
\usepackage{flushend}
\usepackage{soul}

\AtBeginDocument{%
  }

\graphicspath{{figs/}}

\ifx\figurename\undefined \def\figurename{Figure}\fi
\renewcommand{\figurename}{Fig.}
\newcommand{\para}[1]{\textit{\textbf{#1}} }
\renewcommand{\subparagraph}[1]{\underline{\textit{#1}} }

\newcommand{\Sect}[1]{Sec.~\ref{#1}}
\newcommand{\Fig}[1]{Fig.~\ref{#1}}

\newcommand{\Alg}[1]{Algo.~\ref{#1}}

\newcommand{\mode}[1]{\underline{\textsc{#1}}\xspace}

\newcommand{\proj}{\textsc{Atlas}\xspace}

\definecolor{myorange}{RGB}{255, 212, 121}
\newcommand{\circnum}[1]{%
  \tikz[baseline=(char.base)]{
    \node[
      circle,
      draw=black,
      line width=1pt,       
      fill=black,
      inner sep=2pt,
      minimum size=10pt
    ] (outer) {};

    \node[
      circle,
      draw=black,
      line width=0.8pt,     
      fill=orange!50,
      inner sep=1.pt,
      minimum size=8pt
    ] (char) {\sffamily\bfseries #1};
  }%
}

\begin{document}

\title{\proj: Algorithm-Hardware Co-Design for On-Device City-Scale 3D Gaussian Splatting in VR}





\author{He Zhu}
\orcid{0009-0006-0835-7153}
\email{zhcon16@sjtu.edu.cn}

\author{Zheng Liu}
\orcid{0009-0001-6688-4115}
\email{distilledw@sjtu.edu.cn}

\author{Xingyang Li}
\orcid{}
\affiliation{%
  \institution{Shanghai Jiao Tong University}
  \city{Shanghai}
  \country{China}
}
\email{brucelee_sjtu@sjtu.edu.cn}

\author{Anbang Wu}
\orcid{0009-0000-2596-9385}
\email{anbang@cs.sjtu.edu.cn}

\author{Zihan Liu}
\orcid{0000-0002-0874-0682}
\affiliation{%
  \institution{Shanghai Jiao Tong University}
  \city{Shanghai}
  \country{China}
}
\email{altair.liu@sjtu.edu.cn}

\author{Ruyang Li}
\orcid{}
\email{liruyang@ieisystem.com}

\author{Hui Wei}
\orcid{}
\email{weihui@ieisystem.com}

\author{Yaqian Zhao}
\orcid{}
\affiliation{%
  \institution{IEIT SYSTEMS Co., Ltd.}
  \city{Beijing}
  \country{China}
}
\email{zhaoyaqian@ieisystem.com}

\author{Jingwen Leng}
\orcid{0000-0002-5660-5493}
\email{leng-jw@sjtu.edu.cn}

\author{Minyi Guo}
\authornote{Corresponding Authors.}
\orcid{0000-0003-0034-2302}
\email{guo-my@sjtu.edu.cn}

\author{Yu Feng}
\authornotemark[1]
\orcid{0000-0002-2192-5737}
\affiliation{%
  \institution{Shanghai Jiao Tong University}
  \city{Shanghai}
  \country{China}
}
\email{y-feng@sjtu.edu.cn}

\renewcommand{\shortauthors}{He Zhu et al.}

\begin{abstract}

3D Gaussian splatting (3DGS) has drawn significant attention in the architectural community recently.
However, enabling city-scale 3DGS on mobile VR devices remains challenging, as the memory requirement of large-scale scenes far exceeds the memory capacity of today’s mobile GPUs.

This paper presents \proj, an on-device city-scale 3DGS rendering framework that enables scalable rendering without runtime Internet access. 
The key insight is that \textit{although the full 3DGS model is massive, each frame only requires a small subset of Gaussians under the current pose and level-of-detail requirement}. 
Based on this insight, \proj introduces a hierarchical memory offloading mechanism that dynamically loads only necessary Gaussian data into device memory.
To further improve performance, \proj proposes temporal-aware LoD search and stereo rasterization to avoid redundant computation in VR.
We further show that our technique can be integrated with existing 3DGS accelerators with negligible hardware overhead.
Overall, \proj achieves 18.5$\times$ speedup over the GPU baseline and 3.9$\times$ speedup over the state-of-the-art 3DGS accelerators, with 92.4\% energy savings.

\end{abstract}
  


\maketitle

\section{Introduction}
\label{sec:intro}

Neural rendering is ushering in a renaissance in computer graphics by enabling photorealistic and view-dependent rendering, with much higher speeds than conventional ray tracing~\cite{pharr2023physically, deng2017toward, pantaleoni2010hlbvh}.~\footnote{
This paper is an extension of an earlier version, ``\textit{Nebula: Infinite-Scale 3D Gaussian Splatting in VR via Collaborative Rendering and Accelerated Stereo Rasterization}'', that appeared in Proceedings of the 31st ACM International Conference on Architectural Support for Programming Languages and Operating Systems (ASPLOS’26)~\cite{zhu2026nebula}.
We extend the conference version as follows: 

1) we redesign the overall framework to support fully on-device city-scale 3DGS rendering, instead of relying on cloud resources (\Sect{sec:framework}); 

2) we extend temporal-aware tree traversal to support on-device LoD search (\Sect{sec:lod}); 

3) we introduce a hierarchical LoD-tree memory management mechanism to reduce on-device GPU memory pressure (\Sect{sec:mem}); and

4) we expand the evaluation to cover local GPU rendering and the proposed hardware augmentation (\Sect{sec:eval}). 

Compared to the baseline hardware, our unified algorithm-hardware co-design achieves 18.5$\times$ speedup and 13.1$\times$ energy savings, respectively.
} 
In recent years, neural rendering has drawn significant attention in the system community~\cite{feng2025lumina, ye2025gaussian, feng2024potamoi, lee2024gscore, li2025uni, lee2025vr, lin2025metasapiens, durvasula2025arc, he2025gsarch, lee2023neurex, rao2022icarus, li2023instant, mubarik2023hardware, li2022rt, fu2023gen, feng2024cicero, song2024srender, liu2025cambricon}, with 3D Gaussian splatting (3DGS) standing out due to its compact representation and superior rendering performance.

\para{Challenges.} While prior 3DGS acceleration framework~\cite{feng2025lumina, ye2025gaussian, feng2024potamoi, lee2024gscore, li2025uni, lee2025vr, lin2025metasapiens, durvasula2025arc, he2025gsarch, pei2025gcc, oh2026neo, liu2026asdr} achieve real-time mobile rendering for small-scale scenes~\cite{barron2022mipnerf360, hedman2018deep, Knapitsch2017}, they often overlook the scalability challenge of 3DGS, making their designs fragile for large-scale rendering, especially for scenes at the city-scale~\cite{kerbl2024hierarchical, ren2024octree, liu2024citygaussian, li2023matrixcity, wu2025blockgaussian}.
As shown in \Sect{sec:ch:local}, the memory requirement for such scenes can reach up to 66 GB, far exceeding the memory capacity ($<$12~GB) of a typical mobile device in virtual reality (VR)~\cite{questprospec, htcvivespec, visionprospec}.
Moreover, supporting level-of-detail (LoD) further exacerbates the memory pressure, as additional hierarchical representations and LoD-related data must be stored in memory.
This memory gap motivates the need for a new 3DGS rendering framework that drastically reduces on-device memory requirements while preserving rendering performance comparable to systems with effectively unlimited GPU memory.

Despite numerous solutions for cloud-client collaborative rendering~\cite{meng2020coterie, xie2021q, leng2019energy, zhao2020deja, wen2023post0, xu2023edge, zhao2021holoar, zhu2026nebula}, these approaches inherently rely on network connections to offload large 3DGS models or stream intermediate results from remote servers to the host devices.
This dependency makes them less reliable for mobile VR scenarios where network availability cannot always be guaranteed.
Thus, our goal is to enable city-scale 3DGS rendering entirely on a mobile device, without requiring internet access during rendering time.

\para{Insight.}
As mentioned earlier, fully on-device rendering introduces a fundamental memory challenge: city-scale 3DGS scenes require tens of gigabytes of memory, far exceeding the capacity of today’s mobile VR devices. 
To bridge this gap, we propose a hierarchical memory offloading framework, \proj, tailored for on-device city-scale 3DGS rendering. 
Our key insight is that, \textit{although a full city-scale 3DGS model is massive, only a small subset of Gaussians is needed for rendering each frame under the current pose and level-of-detail requirement.} 

Specifically, we exploit two key characteristics of the 3DGS pipeline. 
First, the memory demand is highest in the initial stages, where a large number of Gaussians must be accessed for visibility and LoD selection, but drops sharply in later stages as the pipeline narrows down the active set of Gaussians (\Fig{fig:memory_req}). 
Second, during real-time 3D navigation, the visible Gaussians that are required to be rendered exhibit some spatio-temporal correlations (\Fig{fig:profile_overlap} and \Fig{fig:stereo_overlap}).

\para{Framework.}
By leveraging this insight, we propose \proj, an on-device rendering framework in \Sect{sec:framework} tailored for 3DGS at city scale.
To alleviate the memory pressure, we design a \textit{runtime Gaussian management} system in \Sect{sec:mem} that only retains necessary Gaussians in device memory. 
Specifically, we propose two key contributions. 
First, we introduce a hierarchical offloading strategy that coarsely partitions the entire city-scale scene into small spatial blocks. 
At runtime, the system only loads the blocks on demand that fall within the viewable range under the current camera pose, while keeping the remaining blocks in lower-tier storage, e.g., disks. 
This substantially reduces the amount of scene data that must reside in device memory at any time.
Second, we design an asynchronous runtime Gaussian management mechanism that tracks the active Gaussian working set for each frame and determines which Gaussians should be loaded from disk into device memory on demand.
By overlapping data movement with rendering, our system hides the loading latency and maintains a compact memory footprint without interrupting real-time rendering.

To further boost the overall rendering performance, we introduce two algorithmic optimizations to address the two key bottlenecks, LoD search and rasterization, in the 3DGS rendering pipeline. 
First, we propose a \textit{temporal-aware LoD search} algorithm in \Sect{sec:lod}, which can be directly deployed on existing GPUs.
Our algorithm regularizes DRAM accesses by streaming data during LoD search and exploits temporal coherence across consecutive frames to avoid redundant data accesses and computation.
Second, for VR rendering, where two tightly coupled stereo views must be rendered for left and right eyes, we introduce a \textit{stereo rasterization} pipeline in \Sect{sec:stereo}.
This pipeline leverages triangulation~\cite{hartley2003multiple, szeliski2010computer}, a widely used technique in computer vision, to share most computations across the two eye views while still producing bit-accurate images. 
By reusing shared results, our method skips repetitive stages such as preprocessing and sorting for the second view, while further reducing the cost of rasterization in 3DGS rendering.

\para{Architecture.}
We further show that \proj can be easily integrated into mainstream 3DGS accelerators with minimal hardware augmentation in \Sect{sec:arch}. 
The key component is a lightweight line buffer that enables fine-grained pipelining between left-eye and right-eye rendering in VR.
During left-eye rasterization, many Gaussians that contribute to the left view will also be needed shortly afterward for the right view due to the strong correlation between the two stereo views. 
Instead of fetching these Gaussians again from the on-chip buffer for right-eye rendering, our architecture temporarily keeps the reusable Gaussians in line buffers as they are processed for the left eye. 
The right-eye pipeline can then directly consume these buffered Gaussians from the line buffers, avoiding redundant buffer accesses.

Overall, \proj achieves 18.5$\times$ motion-to-photon speedup and reduces overall energy by 92.4\% compared to a mobile GPU. Even when integrated with state-of-the-art 3DGS accelerators~\cite{lee2024gscore, ye2025gaussian}, \proj achieves 3.9$\times$ speedup and 4.7$\times$ energy savings, while incurring only minimal hardware overhead.

The contributions of this paper are as follows:
\begin{itemize}
    \item We propose an on-device rendering framework tailored for large-scale 3DGS with great scalability.
    \item We propose an asynchronous runtime Gaussian management mechanism that reduces on-device memory pressure by up to 7.0$\times$.
    \item We propose two techniques, temporal-aware LoD search and stereo rasterization, that accelerate two key bottlenecks, LoD search and rasterization, in the 3DGS rendering pipeline by up to 8.1$\times$ and 1.9$\times$.
    \item Our architecture achieves 18.5$\times$ speedup and 7.0$\times$ GPU memory reduction.
\end{itemize}

\section{Large-Scale 3DGS pipeline}
\label{sec:bg}

In this section, we first introduce a key concept in large-scale 3DGS rendering, \textit{hierarchical representation}, and then describe the general rendering pipeline of large-scale 3DGS algorithms.

\para{Hierarchical Representations.} 
Compared to small-scale 3DGS algorithms~\cite{kerbl20233d, fan2023lightgaussian, fang2024mini, mallick2024taming, feng2024flashgs, gui2024balanced, zhu2026seele, wang2024adr}, large-scale 3DGS algorithms~\cite{wu2025blockgaussian, kerbl2024hierarchical, liu2024citygaussian, ren2024octree} introduce hierarchical representations to manage the vast number of Gaussian ellipsoids, i.e., the fundamental rendering primitives in 3DGS. 
These hierarchical representations enable LoD rendering through LoD search, avoiding unnecessary computation for objects at a far distance, similar to the role of mipmaps in traditional rasterization pipelines~\cite{kilgard2000practical, akenine2019real}. 

During global view rendering, e.g., bird eye views, LoD allows the pipeline to render large regions at a coarser granularity, since rendering fine details introduces additional computational overhead without noticeable quality improvement. 
In this case, multiple small Gaussians far from the camera can be represented by a single larger Gaussian for rendering. 
In contrast, during local view rendering, e.g., navigating street blocks, LoD naturally follows a ``divide-and-conquer'' strategy through its hierarchical representation: nearby regions are rendered with fine details, while distant regions are more coarse. 
Moreover, LoD search can efficiently cull irrelevant Gaussians that fall outside the current view frustum or are far away from the current location, thereby reducing the overall rendering workload.

\begin{figure*}[t]
    \centering
    \includegraphics[width=0.95\textwidth]{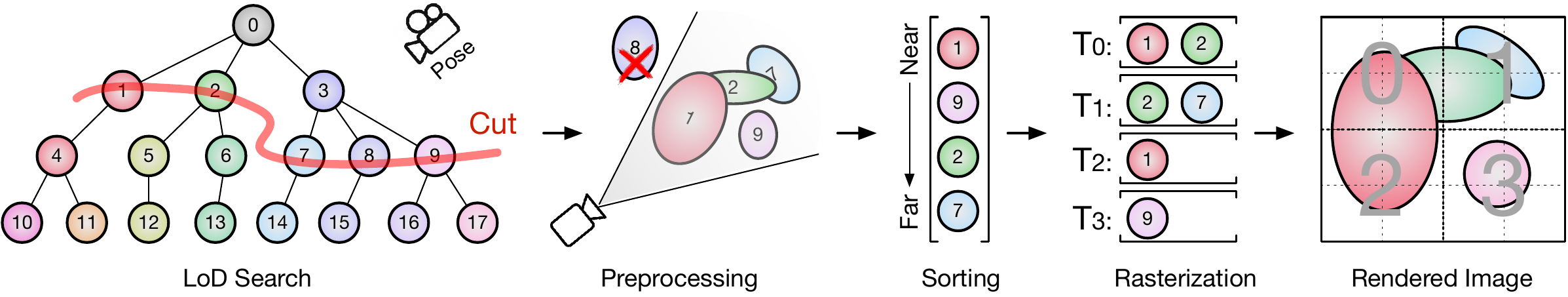}
    \caption{The rendering pipeline for large-scale 3DGS consists of four stages: LoD search, preprocessing, sorting, and rasterization.
    First, LoD search traverses the LoD tree to determine a set of Gaussians with a desired LoD granularity.
    The result Gaussians form a ``cut'' that separates the top and bottom of the LOD tree.
    Then, the Gaussians on the cut go through a sequence of operations, i.e., preprocessing, sorting, and rasterization, to render an image, similar to the small-scale 3DGS pipelines~\cite{kerbl20233d}.}
    \label{fig:pipeline}
\end{figure*}

\para{LoD Tree.}
The left part of \Fig{fig:pipeline} illustrates an example hierarchical representation used to organize all Gaussians, referred to as the \textit{LoD tree}, where each tree level corresponds to a specific level of detail. Each tree node contains a single Gaussian, and its child nodes represent finer-grained details associated with the parent node. A LoD tree can be implemented using various tree-like structures, such as an octree~\cite{ren2024octree}, an irregular tree~\cite{kerbl2024hierarchical}, or a shallow tree where each leaf node stores a flattened list of Gaussians~\cite{liu2024citygaussian, wu2025blockgaussian}.

In this paper, we describe the most general form of the LoD tree: an irregular tree in which each node represents one Gaussian and may contain an arbitrary number of child nodes. Gaussians at lower levels of the tree generally capture finer details. Other tree-like structures can be viewed as special cases of this general representation.

\para{Pipeline.} 
As shown in \Fig{fig:pipeline}, a general large-scale 3DGS rendering pipeline consists of four main stages: \textit{LoD search}, \textit{preprocessing}, \textit{sorting}, and \textit{rasterization}. 

\subparagraph{LoD Search.} 
This stage selects Gaussians at an appropriate LoD for the subsequent rendering stages. Specifically, the renderer traverses the LoD tree from top to bottom. 
At each node, it checks whether the projected size of the Gaussian is smaller than a predefined LoD threshold $\tau^*$, i.e., a target pixel-space size, while the projected size of its parent node is larger than the threshold at the current camera pose. 
Gaussians satisfying this condition are selected for rendering.
Conceptually, these selected Gaussians form a ``cut'' that separates the upper and lower parts of the LoD tree.

\subparagraph{Preprocessing.}
Once the cut is determined, the selected Gaussians are projected onto the rendering canvas. Gaussians outside the view frustum, e.g., Gaussian 8 in \Fig{fig:pipeline}, are filtered out.

\subparagraph{Sorting.}
The remaining Gaussians are sorted by depth, from nearest to farthest, to ensure visually consistent blending.

\subparagraph{Rasterization.}
The final stage blends the sorted Gaussians onto the image. 
This process is performed tile by tile. 
Each tile first identifies the Gaussians that intersect with it and constructs a local Gaussian list, as shown in \Fig{fig:pipeline}. 
For example, tile $T_0$ only intersects Gaussians 1 and 2. 
Next, each pixel within the tile performs \textit{$\alpha$-checking}, which computes the intersected transparency $\alpha_i$ for each overlapping Gaussian. 
If $\alpha_i$ is below a predefined threshold, the pixel skips this Gaussian during color blending. 
Otherwise, the Gaussian contributes to the final pixel color through weighted blending. 
The pixel color is then obtained from its contributing Gaussians based on,
\begin{align}
\label{eqn:nerf}
   C(\textbf{p}) & = \sum_{i=1}^{N} \Gamma_i \alpha_i \textbf{c}_{i},\quad \text{and} \quad \Gamma_i = \prod^{i-1}_{j=1} (1-\alpha_j),
\end{align}
where $C(\mathbf{p})$ denotes the final color of pixel $\mathbf{p}$, $\alpha_i$ and $\mathbf{c}_i$ represent the transparency and color of the $i$th Gaussian, and $\Gamma_i$ is the accumulated transmittance along the ray from the first Gaussian to the $(i-1)$th Gaussian. 
The transparency $\alpha_i$ of each Gaussian is determined by its opacity $\sigma_i$ and its 2D covariance matrix~\cite{zwicker2001ewa}.

\section{Challenges and Opportunities}
\label{sec:ch}

We first describe the challenges in large-scale 3DGS under local rendering (\Sect{sec:ch:local}).
We then explain the insights that can be exploited to address those challenges (\Sect{sec:ch:op}).

\subsection{Challenges in Local Rendering}
\label{sec:ch:local}

\begin{figure}[t]
\centering
\begin{minipage}[t]{0.68\columnwidth}
    \centering
    \includegraphics[width=\columnwidth]{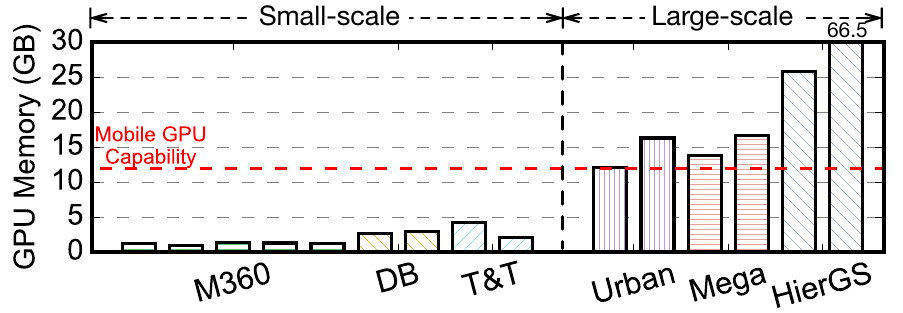}
    \caption{GPU memory footprint trends with scene scale. Runtime numbers are measured across six datasets. Small-scale datasets: T\&T~\cite{Knapitsch2017}, DB~\cite{hedman2018deep}, and M360~\cite{barron2022mipnerf360}. Large-scale datasets: Urban~\cite{lin2022capturing}, Mega~\cite{turki2022mega}, and HierGS~\cite{kerbl2024hierarchical}.}
    \label{fig:memory_pressure}
\end{minipage}
\hspace{2pt}
\begin{minipage}[t]{0.3\columnwidth}
  \centering
  \includegraphics[width=\columnwidth]{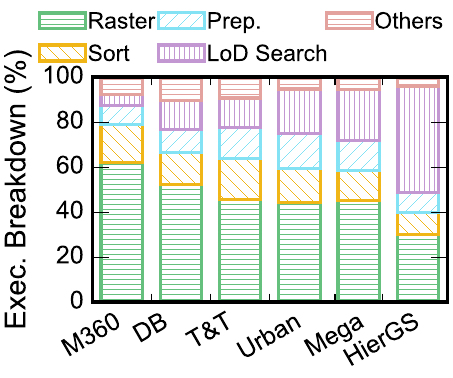}
  \caption{The end-to-end execution breakdown of \textit{local rendering} on a mobile Ampere GPU~\cite{orinsoc}. ``Others'': time on sensor tracking and display.}
  \label{fig:exec_time}
\end{minipage}
\end{figure}

\para{Memory Pressure.}
The first challenge in local rendering is the memory pressure imposed by the massive scale of large-scene 3DGS models.
\Fig{fig:memory_pressure} shows the runtime GPU memory footprint across scenes in different datasets~\cite{barron2022mipnerf360, hedman2018deep, Knapitsch2017, kerbl2024hierarchical, lin2022capturing, turki2022mega}.
As the rendering scene scales from small to large, memory usage grows drastically and quickly drains the capacity of mobile GPUs.
All scenes from large-scale datasets exceed the memory capacity of mainstream VR devices~\cite{questprospec, quest3spec, htcvivespec, visionprospec}, which are often less than 12 GB.
Specifically, a scene from HierGS~\cite{li2023matrixcity} even exceeds 66~GB.
However, prior architectural designs~\cite{feng2025lumina, ye2025gaussian, feng2024potamoi, lee2024gscore, li2025uni, lee2025vr, lin2025metasapiens, durvasula2025arc, he2025gsarch} have primarily focused on small scenes, largely ignoring the scalability challenges posed by large-scale 3DGS contents.
Without addressing this bottleneck, it is infeasible to achieve infinite-scale 3DGS rendering in the foreseeable future.

\para{Bottleneck Shift.}
Another key observation in large-scale 3DGS rendering is that, as the scene complexity increases, the computational bottleneck shifts from \textit{rasterization} to \textit{LoD search}. 
Quite a few studies~\cite{lee2024gscore, feng2025lumina, ye2025gaussian, lin2025metasapiens} propose dedicated accelerators for rasterization, which dominates the execution time in small-scale scenes in \Fig{fig:exec_time}. 
However, with increasing scene complexity, the cost of LoD search, i.e., identifying which Gaussians should be rendered, increases rapidly and begins to dominate the overall execution.

As shown in \Fig{fig:exec_time}, the relative execution time of LoD search increases with the scene size on a Nvidia mobile Ampere GPU~\cite{orinsoc}, accounting for up to 47\% of the end-to-end latency in large-scale 3DGS scenes.
In contrast, the relative time of rasterization does not grow with scene scales, because, with LoD search, the number of Gaussians that can contribute to the final frame plateaus. 
Therefore, without effectively supporting LoD search, it remains infeasible to achieve real-time rendering on mobile devices.

\begin{figure}[t]
\centering
\begin{minipage}[t]{0.32\columnwidth}
  \centering
  \includegraphics[width=\columnwidth]{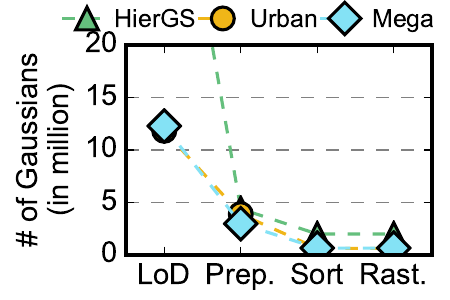}
  \caption{The runtime memory demand varies across different stages. We use the number of involved Gaussians as a proxy for memory demand.}
  \label{fig:memory_req}
\end{minipage}
\hspace{2pt}
\begin{minipage}[t]{0.32\columnwidth}
  \centering
  \includegraphics[width=\columnwidth]{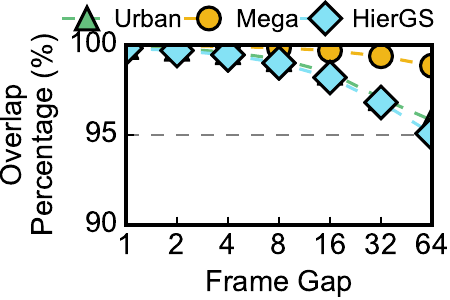}
  \caption{The temporal similarity between adjacent frames under a 90~FPS VR scenario.}
  \label{fig:profile_overlap}
\end{minipage}
\hspace{2pt}
\begin{minipage}[t]{0.32\columnwidth}
  \centering
  \includegraphics[width=\columnwidth]{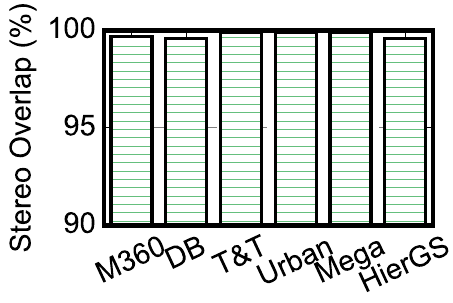}
  \caption{The stereo similarity between the left-eye and right-eye images in VR.}
  \label{fig:stereo_overlap}
\end{minipage}
\end{figure}

\subsection{Key Insights}
\label{sec:ch:op}

\para{Memory Demand.}
Although \Sect{sec:ch:local} shows that large-scale 3DGS scenes exceed the memory capacity of VR devices, we observe that the runtime memory demand varies substantially across different stages of the 3DGS pipeline. 
\Fig{fig:memory_req} quantifies this variation by using the number of rendered Gaussians as a proxy for memory demand.

The initial stage requires the highest memory, as LoD search may need to examine a large portion of the scene to determine the appropriate level of detail. 
After the LoD search, however, the number of active Gaussians drops rapidly to a scale that can be accommodated by a mobile device.
This result shows that the 3DGS pipeline can be naturally divided into two parts: a high-memory-demand LoD search stage and a lower-memory rendering stage. 
Therefore, carefully managing the memory-intensive LoD search stage is essential for enabling large-scale 3DGS rendering under a memory-constrained mobile device.

\para{Temporal Similarity.}
Meanwhile, we find that the set of Gaussians selected by LoD search, i.e., the ``cut'' in \Fig{fig:pipeline}, exhibits strong temporal similarity across adjacent frames. 
\Fig{fig:profile_overlap} shows the overlap ratio of selected Gaussians after LoD search on the HierGS dataset~\cite{kerbl2024hierarchical}.
Here, we simulate a 90 FPS VR rendering scenario. 
The results show that over 99\% of the Gaussians selected after LoD search remain unchanged between two consecutive frames. 
Even when the frame gap exceeds 64, more than 95\% of the selected Gaussians remain identical.
This high temporal similarity shows that a substantial portion of LoD search computation is redundant across frames.

\para{Stereo Similarity.}
In addition to temporal similarity across frames, the left- and right-eye views in VR also exhibit strong stereo similarity.
\Fig{fig:stereo_overlap} shows the percentage of overlapping pixels between the two views across different datasets. 
To quantify this overlap, we warp the left-eye image to the right-eye view using a technique similar to Cicero~\cite{feng2024cicero}. 
Our results show that fewer than 1\% of pixels are non-overlapping between the left and right views.

However, we cannot directly reuse the warped left-eye pixels for the right eye.
Because 3DGS rendering is \textit{view-dependent}, it means that the same physical point may appear with different colors under different viewing directions, e.g., due to specular reflection.
Therefore, directly warping pixels from the left eye to the right eye can introduce noticeable artifacts.
This motivates us to design a stereo-aware rendering strategy that reuses computation.

\para{Summary.}
Overall, we show three key opportunities for optimizing on-device city-scale 3DGS rendering. 
\begin{itemize}
    \item First, memory demand is highly stage-dependent: while LoD search requires access to a large portion of the scene, the active Gaussian set after LoD search becomes small enough for mobile GPUs.
    \item Second, LoD search results exhibit strong temporal similarity across consecutive frames.
    Thus, much of the search computation can be reused rather than recomputed from scratch. 
    \item Third, VR stereo rendering shows substantial overlap between the left- and right-eye views.
    It shows opportunities to share computation across the two rendering pipelines.
\end{itemize}

\section{\proj Overview}
\label{sec:framework}

To address the challenges in \Sect{sec:ch}, we introduce our rendering framework, \proj.
In this section, we first give an overview of three key contributions in \proj.
We then explain our temporal-aware LoD search in \Sect{sec:lod}.
Next, we show how to manage massive Gaussian data on a mobile device in \Sect{sec:mem}.
Lastly, we introduce our novel stereo rendering pipeline in \Sect{sec:stereo}.

\begin{figure*}[t]
    \centering
    \includegraphics[width=\textwidth]{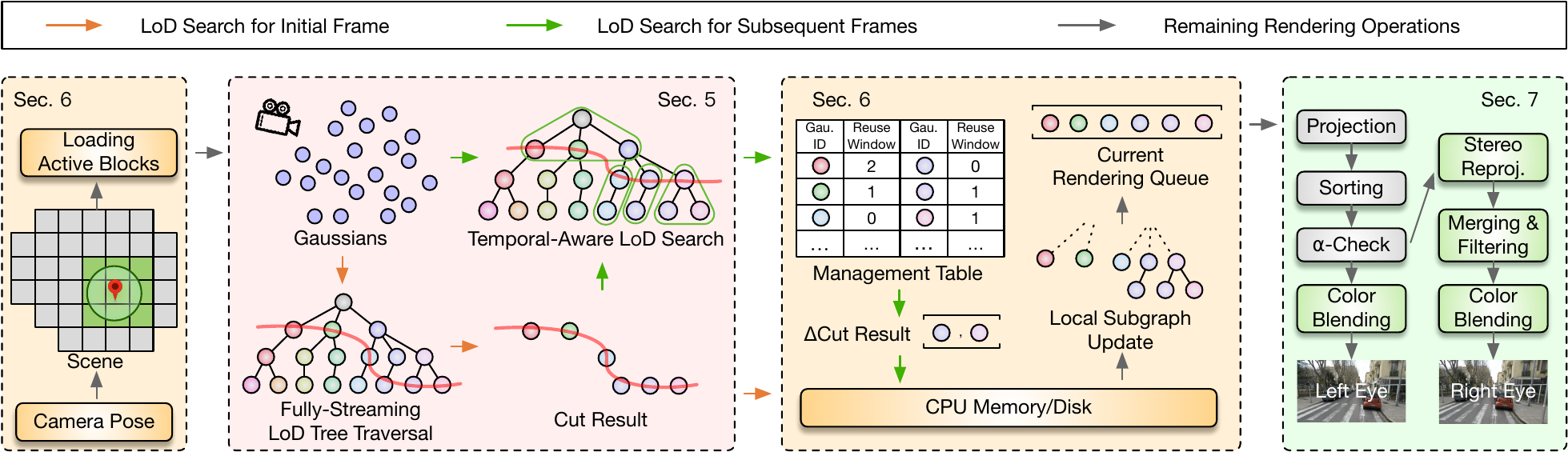}
    \caption{Overview of \proj workflow. \proj enables fully on-device city-scale 3DGS rendering via hierarchical memory management. 
    The scene is partitioned into spatial blocks, and only the lightweight geometric metadata of active blocks is kept in GPU memory for local LoD search. 
    Rendering-related Gaussian attributes are loaded from lower-tier storage on demand.
    \proj combines temporal-aware LoD search (\Sect{sec:lod}), runtime Gaussian management for loading and eviction (\Sect{sec:mem}), and stereo rasterization to reduce redundant computation across two eye views (\Sect{sec:stereo}).}
    \label{fig:overview}
\end{figure*}

\para{Idea.}
Guided by our design principles, \proj exploits the three key insights in \Sect{sec:ch:op}.
First, to alleviate the memory pressure of LoD search in large-scale 3DGS, instead of maintaining all Gaussian attributes in GPU memory, we keep only the necessary metadata required for LoD search, i.e., positions and scales, in device memory.
For the remaining Gaussian attributes, we offload them to lower-tier memory, e.g., CPU memory or disk.
This allows \proj to execute LoD search locally without requiring the full Gaussian model to reside in GPU memory.
Second, once LoD search identifies the subset of Gaussians required for rendering, \proj loads their remaining attributes on demand from lower-tier storage into GPU memory. By exploiting temporal similarity across frames, our runtime Gaussian management system avoids repeatedly loading unchanged Gaussians across frames and only fetches newly required attributes.
Lastly, our rendering pipeline exploits the stereo similarity of the binocular view to avoid redundant computation across two eyes.

Note that \textit{\proj does not rely on cloud resources or runtime internet connectivity}.
Instead, all computation is performed on the local device, while the large-scale 3DGS model is managed through our hierarchical Gaussian management system.
This design is particularly important for VR scenarios, where the network cannot always be available.

\para{Workflow.}
\Fig{fig:overview} shows the overall workflow of our pipeline. Unlike cloud-client collaborative frameworks, \proj performs LoD search locally on the device. 
To make this feasible under limited GPU memory, \proj first partitions the entire city-scale scene into small spatial blocks.
At runtime, only the blocks near the current view pose are activated and loaded into GPU memory; the remaining blocks stay in lower-tier storage and are fetched on demand when the user moves toward new regions.

Within each active block, Gaussian data is further separated into two categories. 
The first category contains lightweight LoD metadata, i.e., position and scale, which is required for LoD search. 
The second category contains storage-intensive rendering attributes, such as spherical harmonic (SH) coefficients, which are needed only after the LoD search.
Therefore, \proj retains only the LoD metadata of active blocks in GPU memory, while offloading both inactive blocks and remaining rendering attributes to CPU memory or disk storage.

At the beginning of the rendering pipeline, the local device executes LoD search using the LoD metadata of the active blocks.
During this process, we also propose a GPU-efficient \textit{temporal-aware LoD search} in \Sect{sec:lod} that leverages temporal similarity across frames to avoid unnecessary tree node accesses. 
Specifically, \proj processes the initial frame and subsequent frames using different LoD search algorithms.

For the initial frame, \proj performs a full LoD search, i.e., a LoD tree traversal, locally on the device to find the Gaussians at the appropriate LoD, referred to as the ``cut'' (\Fig{fig:pipeline}), under the current pose. 
To avoid irregular memory accesses commonly existing in tree traversal~\cite{pinkham2020quicknn, feng2022crescent, xu2019tigris, feng2020mesorasi, feng2025streamgrid}, we propose a \textit{fully streaming LoD tree traversal} that regularizes memory accesses and is highly parallelizable on GPUs. 
If the current view is close to the boundary of active blocks, \proj asynchronously loads the LoD metadata of neighboring blocks so that the data of subsequent LoD searches are available in GPU memory without stalling the rendering pipeline.

For subsequent frames, \proj leverages the temporal similarity of the cut results across adjacent frames. 
Instead of performing a complete tree traversal from the root node, our temporal-aware LoD search performs lightweight local updates by searching only within relevant subtrees in the active blocks to update the cut result of the current frame. 
Further explanations of our algorithm are provided in \Sect{sec:lod}.

Once the cut result is obtained, \proj determines which selected Gaussians already have their rendering attributes available in GPU memory and which ones are missing. 
Our \textit{runtime Gaussian management} system in \Sect{sec:mem} maintains this information at runtime. 
For missing Gaussians, the system asynchronously loads their remaining attributes from CPU memory or disk storage into GPU memory on demand. 
For Gaussians that are no longer needed, the system evicts their rendering attributes from GPU memory to release GPU memory pressure. 
Similarly, when spatial blocks become irrelevant to the current view pose, their LoD metadata can also be evicted from GPU memory.
By exploiting both spatial locality across blocks and temporal similarity across frames, \proj keeps the GPU memory footprint compact while ensuring that the required Gaussians are available for rendering.

\begin{figure*}[t]
    \centering
    \includegraphics[width=\textwidth]{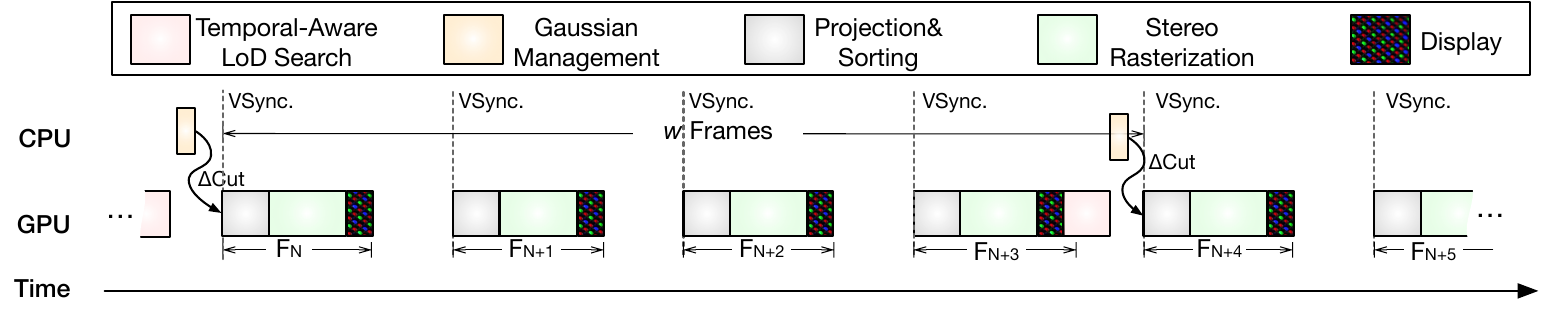}
    \caption{The timing diagram of our execution flow. LoD search is executed locally, while inactive blocks and storage-intensive Gaussian attributes are asynchronously loaded from lower-tier memory on demand. Only rendering operations are on the critical path.}
    \label{fig:exec_flow}
\end{figure*}

After the required Gaussian attributes become available, the device updates its local active Gaussian set and executes the remaining rendering stages.
Lastly, our \textit{stereo rasterization} pipeline uses the active Gaussians in GPU memory to render the left- and right-eye images. 
Instead of rendering two independent frames for the two eyes, our pipeline exploits the geometric relationship between Gaussians and the stereo camera. 
This allows the two views to share most computations while still producing bit-accurate images as if they were rendered separately. 
We further explain our stereo rasterization algorithm in \Sect{sec:stereo}.

\para{Timing.}
\Fig{fig:exec_flow} summarizes the timing of our execution flow. On the local device, \proj maintains that the LoD metadata of blocks near the current view pose is available in GPU memory. 
It then performs temporal-aware LoD search.
Once we obtain the ``cut'' result, the runtime Gaussian management system checks the resulting cut and identifies the Gaussians whose rendering attributes are missing from GPU memory.
These missing attributes are asynchronously loaded from CPU memory or disk storage into GPU memory. 
Meanwhile, the renderer continues processing frames using the active Gaussian set already available in GPU memory.

Following the previous large-scale 3DGS pipeline~\cite{kerbl2024hierarchical}, LoD search is executed only once every $w$ frames, e.g., $w = 4$. 
Between two LoD searches, \proj reuses the active Gaussian set and performs local rendering updates. 
Since adjacent frames are temporally similar, the number of newly required Gaussian attributes is negligible. 
As a result, most memory traffic can be hidden behind rendering. 
Once rendering completes, the new image is displayed at the next VSync arrival.

\section{Temporal-Aware LoD Search}
\label{sec:lod}

In this section, we first describe how we partition the original LoD tree into subtrees to accommodate our proposed algorithm. 
We then introduce our fully-streaming tree traversal algorithm to accelerate LoD search. 
Finally, we show our temporal-aware LoD search, which leverages previous cut results to avoid redundant tree traversal.

\subsection{LoD Tree Partitioning}
\label{sec:lod:tree}

\begin{algorithm}[t]
\caption{Algorithm of Subtree Partitioning}
\label{algo:subtree}
\KwData{LoD tree $ \{n\} $, subtree size limit $N_{size}$ }
\KwResult{subtrees $ S_{final} $}
$R \leftarrow \{n\}.\text{findRoot()}$, $S_{init}\leftarrow \{\ \}$\;
\While{$R$ is not empty}{
  $n_r \gets R.\text{dequeue()}$\;
  $s_j, \{n_{child}\} \gets \text{BFS(}n_r, \{n\}, N_{size}\text{)}$\;
   $S_{init}\text{.push(}s_j\text{)}$\;
  \For{ $n_i$ in $\{n_{child}\}$ }{
    $ R\text{.enqueue(} n_i \text{)} $\;
  }
}

$S_{final}\leftarrow \{\ \}$\;
$s_{merge} \leftarrow \text{first}(S_{init})$, $S_{init}\leftarrow S_{init} \setminus \{s_{merge}\}$\;

\For{$s \in S_{init}$}{
    \If{ $s.$\text{parent()} is $s_{merge}$.\text{parent()} \textbf{and} $s.\text{size()} \le \tau_s/2$ \textbf{and}
     $s.\text{size()} + s_{merge}.\text{size()} \le \tau_s$}{
        $s_{merge}\leftarrow \text{merge(}s_{merge}, s\text{)}$\;
    }
    \Else{
        $S_{final}\text{.push(}s_{merge}\text{)}$\;
        $s_{merge} \leftarrow s$\;
    }
}
$S_{final}\text{.push(}s_{merge}\text{)}$\;
\end{algorithm}

The process of subtree partitioning is described as follows.

Our subtree partitioning begins with a breadth-first search (BFS) traversal from the root of the LoD tree, grouping nodes progressively as described in \Alg{algo:subtree}. 
During the traversal, each tree node is inserted into the current subtree, $s_j$, until the cumulative number of traversed nodes exceeds the predefined subtree size limit, $N_{size}$. 
At that point, the collected nodes are grouped into a subtree, $s_j$, and then identify the immediate child nodes ${n_{child}}$ of $s_j$. 
These nodes serve as the new roots of their corresponding LoD subtrees and are enqueued in $R$ (as shown in \Alg{algo:subtree}).
The BFS traversal is then applied individually to each new root in $R$. 
This process of partitioning continues recursively until all nodes in the original LoD tree are assigned to subtrees.

Although the partitioning method above splits the original LoD tree into subtrees, some subtrees can be very small (e.g., a single node as a subtree), which may cause workload imbalance across GPU warps during traversal. 
To mitigate this, we introduce a subtree merging process to reduce size variation among subtrees. 
Our key observation is that certain subtrees can be merged without violating the hierarchical relationships of the LoD tree. 
Specifically, we iterate through the initially partitioned subtrees, identify small ones (size < $N_{size}/2$), and attempt to merge small subtrees if they share the same parent. 
The merging is performed greedily and stops once the combined subtree $s_{merge}$ exceeds the size threshold $N_{size}$. 
After subtree merging, we can generate a more balanced subtree structure.

\subsection{Temporal-Aware LoD Tree Traversal}
\label{sec:lod:search}

\begin{figure}[t]
    \centering
    \includegraphics[width=\columnwidth]{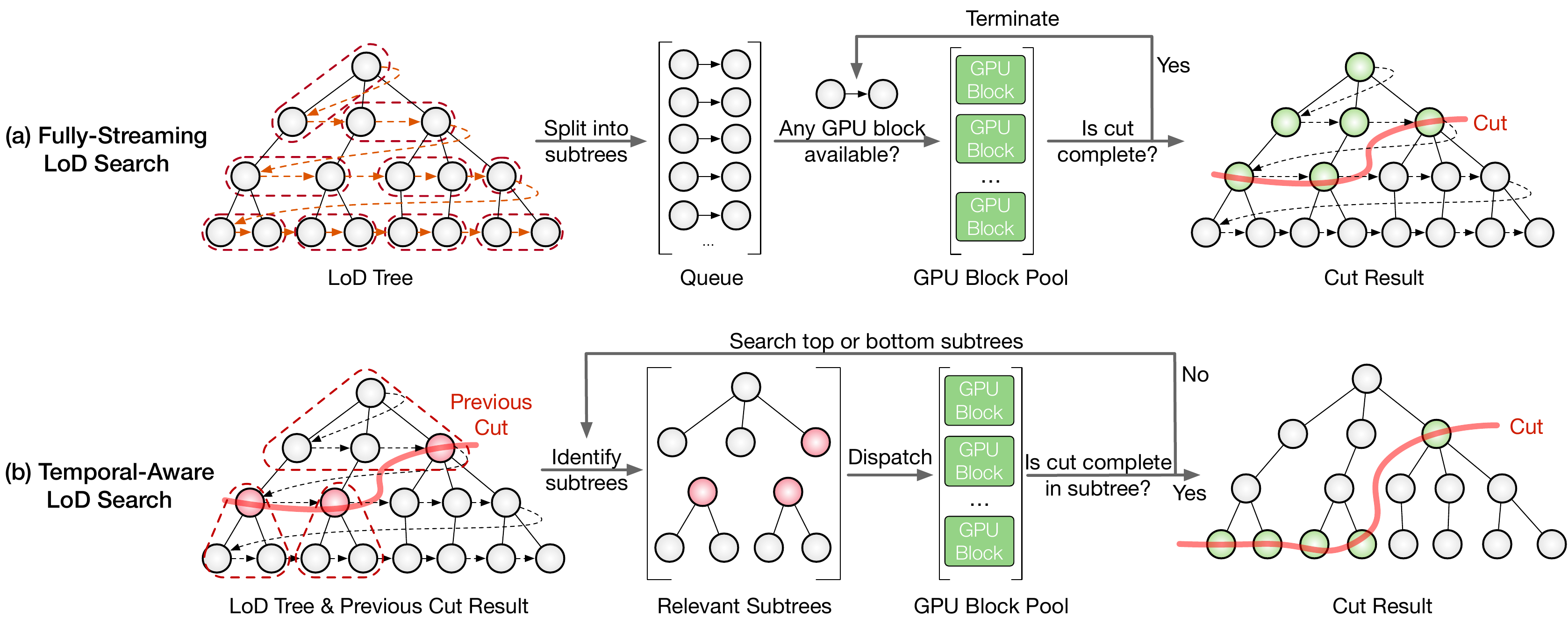}
    \caption{Illustration of fully-streaming LoD tree traversal and temporal-aware LoD search under block-based scene partitioning.}
    \label{fig:tree_traversal}
\end{figure} 

\para{Fully-Streaming LoD Tree Traversal.}
\Fig{fig:tree_traversal}a shows our \textit{fully-streaming LoD tree traversal} when the current view enters a region whose LoD tree has not been searched before.
Recall that \proj partitions the city-scale scene into multiple spatial blocks and keeps only the LoD metadata of nearby active blocks in GPU memory. 
Therefore, LoD search is performed locally over the LoD trees of active blocks rather than over the entire city-scale model.

Within each active block, the LoD tree is an irregular tree where each node represents one Gaussian and may have an arbitrary number of child nodes. 
A conventional tree traversal over such an irregular tree leads to irregular memory accesses and poor GPU utilization~\cite{pinkham2020quicknn, feng2022crescent}.
Our goal is to achieve high GPU parallelism while minimizing unnecessary tree node visits and keeping the memory footprint small.
Rather than relying solely on the inherent parent-child relations in the original LoD tree (denoted by solid arrows), we augment each block-local LoD tree with additional connections that enable traversal in BFS order (denoted by orange dashed arrows).

During GPU execution, each GPU warp is assigned an equal-sized workload, i.e., a block of tree nodes from the currently active spatial blocks.
Each block of tree nodes is then evenly distributed among threads within a warp to ensure a balanced workload across GPU threads. 
We design each node block to be small enough to reside entirely in GPU shared memory, allowing streaming access to each tree block and avoiding irregular DRAM accesses.
Workload assignment is dynamically dispatched whenever a GPU warp becomes available at runtime.
The traversal of tree blocks terminates once a clean cut separates the upper and lower parts of its LoD tree (see the red curve in \Fig{fig:tree_traversal}a).
In this way, our algorithm visits only the necessary nodes, e.g., the green nodes in \Fig{fig:tree_traversal}a, while skipping irrelevant nodes, e.g., the grey nodes, thereby reducing both redundant computation and memory traffic.

Importantly, this traversal only requires lightweight LoD metadata, i.e., position and scale, which is resident for active spatial blocks.
Storage-intensive rendering attributes are not loaded during LoD search. 
They are fetched later on demand only for Gaussians selected by the cut, as described in \Sect{sec:mem}.
This separation allows \proj to execute LoD search locally while keeping GPU memory usage tight.

\para{Temporal-Aware LoD Search.}
For subsequent frames, we introduce a \textit{temporal-aware LoD search} that exploits temporal similarity across frames. 
During offline, each block-local LoD tree is partitioned into multiple subtrees while preserving its hierarchical relationships.
\Fig{fig:tree_traversal}b highlights these subtrees using dashed blocks. 
For illustration, we show a two-level subtree partitioning, while our implementation supports multi-level partitioning.

At runtime, given the previous frame's cut result (highlighted in pink), our algorithm identifies the subtrees that contain the Gaussians in the previous cut. 
\Fig{fig:tree_traversal}b highlights these subtrees with red dashed blocks. 
For the current frame, \proj first searches only these identified subtrees, instead of traversing all subtrees in all active blocks. 
In our GPU implementation, each subtree is assigned to a separate GPU warp for local traversal. 
Since subtree partitioning is performed offline and each subtree is constructed to have approximately equal size, this design ensures balanced workload distribution across GPU warps.

When the user moves or rotates the headset, the set of active spatial blocks may change. 
If the current pose remains within the previously active region, temporal-aware LoD search updates the cut by searching only the relevant subtrees inside the already active blocks. 
If the pose approaches a block boundary or exposes a previously inactive region, \proj asynchronously activates the neighboring blocks and loads their lightweight LoD metadata from lower-tier storage. 
For newly activated blocks that do not have a previous cut, \proj applies the fully-streaming LoD traversal described above to compute their initial cut. 
Thus, temporal-aware search handles stable regions efficiently, while fully-streaming traversal handles newly activated blocks.

If searching the local subtree cannot produce a complete cut, i.e., the subtree does not contain a clean boundary for the current LoD threshold, the algorithm expands the search to its corresponding parent subtree or neighboring subtrees until the cut is completed. 
This fallback preserves correctness while still avoiding full-tree traversal in most frames. 
The resulting cut is then passed to the runtime Gaussian management system, which determines which Gaussians already have their rendering attributes in GPU memory and which attributes must be loaded on demand from disk.

Note that our temporal-aware LoD search produces \textit{bit-accurate} results compared to the original full-tree traversal. In \Fig{fig:lod_speedup}, we compare our tree traversal performance against prior designs and achieve up to 8.1$\times$ speedup.

\section{Runtime Gaussian Management}
\label{sec:mem}

We next describe how \proj manages Gaussians at runtime to enable fully on-device city-scale 3DGS rendering under limited GPU memory.
Unlike cloud-client collaborative rendering in \textsc{Nebula}~\cite{zhu2026nebula}, \proj does not offload massive Gaussians to the cloud. 
Instead, it dynamically moves Gaussian data within its local memory hierarchy, i.e., between GPU memory, CPU memory, and disk storage. 
Our runtime Gaussian management system must guarantee two key properties: 1) only the Gaussians required by the current and near-future frames are kept in GPU memory; and 2) obsolete Gaussians are removed on the fly to alleviate memory pressure.
We now describe how \proj manages spatial blocks and Gaussian attributes at runtime.

\begin{figure}[t]
    \centering
    \includegraphics[width=\columnwidth]{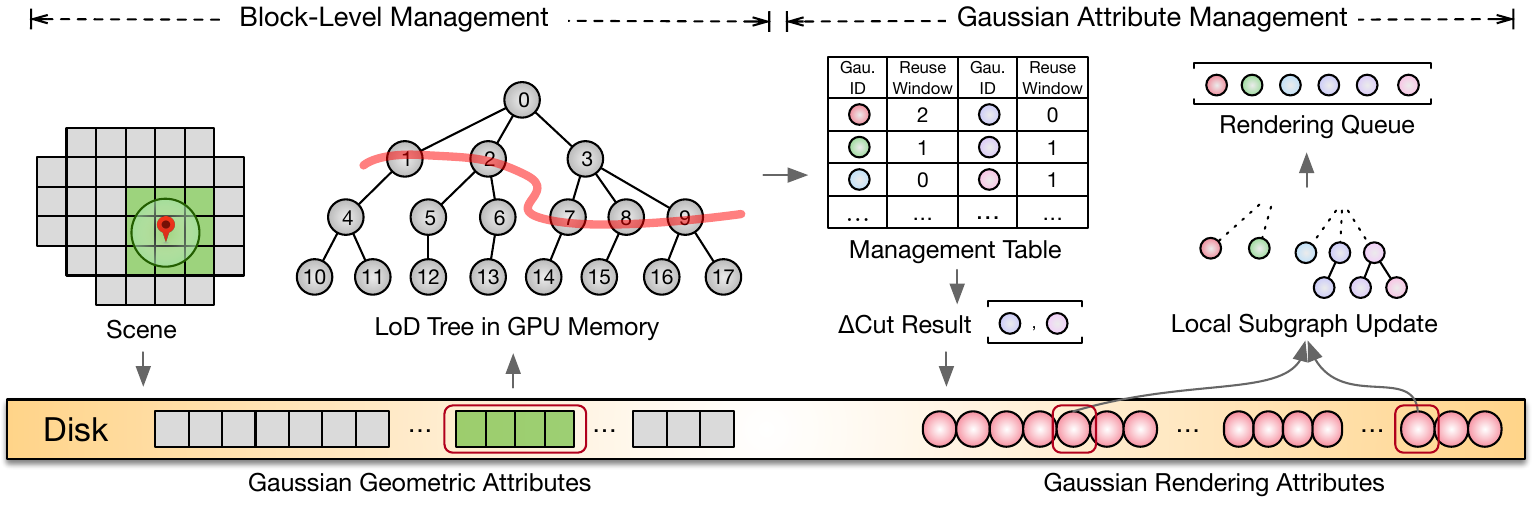}
    \caption{Illustration of our runtime Gaussian management system.}
    \label{fig:gaussian_management}
\end{figure} 

\para{Block-Level Management.}
As shown in \Fig{fig:gaussian_management}, \proj first partitions the entire city-scale scene into multiple spatial blocks. 
Each block contains a subset of the LoD tree and its associated Gaussians. 
At runtime, \proj maintains an \textit{active block set} based on the current camera pose and viewable range.
For all Gaussians in the active block set, only the lightweight LoD attributes, i.e., positions, scales, and tree connectivity, are loaded into GPU memory for local LoD search.
All metadata of inactive blocks remains in lower-tier storage, such as CPU memory or disk.

When the current pose approaches the boundary of active blocks, \proj proactively activates neighboring blocks and asynchronously loads their LoD metadata into GPU memory.
Conversely, blocks that have not been used for a certain number of frames are marked as inactive and evicted from GPU memory.
This block-level management ensures that LoD search can still be executed locally while avoiding keeping the entire LoD tree in GPU memory.

\para{Gaussian Attribute Management.}
Within each active block, there is a small LoD tree. 
\proj separates the Gaussian attributes in this tree into two categories.
The first category contains lightweight LoD metadata required for LoD search.
These data are always kept in the GPU memory.
The second category contains rendering attributes, such as SH coefficients, which are only needed after the LoD search determines which Gaussians should be rendered.

To manage the second category, \proj maintains a runtime Gaussian table that tracks which Gaussians currently have their rendering attributes kept in GPU memory. 
For each Gaussian, this runtime Gaussian table records its residency status and a reuse window, denoted as $w_r$, which counts the number of rendering frames since this Gaussian was last included in a cut result and is reset whenever the Gaussian is selected by a new cut.
When a new cut result is generated by LoD search, \proj checks the Gaussian table and identifies Gaussians whose rendering attributes are missing from GPU memory. 
These missing attributes are then gathered into a request group and asynchronously loaded from CPU memory or disk storage into GPU memory.

Meanwhile, \proj also removes obsolete Gaussians to keep GPU memory usage compact.
We use a reuse threshold $w_r^*$, which is set to 32 in our implementation. 
After each table update, \proj scans the Gaussian table and evicts any Gaussian whose reuse window $w_r$ exceeds $w_r^*$.
These Gaussians are considered unlikely to be reused in the near future, and their rendering attributes are removed from GPU memory.
The overall idea is similar to garbage collection~\cite{lieberman1983real, appel1989simple}, but is specialized for the 3DGS rendering.

\para{Local Rendering Queue Generation.}
After the required Gaussian attributes are loaded, \proj updates the local active Gaussian set, which only stores the Gaussians associated with active blocks and recently used cut results.
Each Gaussian entry contains its metadata, residency status, and pointers to rendering attributes when available.

Lastly, \proj generates a rendering queue containing Gaussians at the appropriate LoD and submits to the rasterization pipeline.
Only Gaussians whose rendering attributes are available in GPU memory are inserted into the rendering queue. 
By combining block-level activation, on-demand attribute loading, and reuse-window-based eviction, \proj keeps GPU memory usage low while enabling fully on-device rendering of city-scale 3DGS scenes.

\section{Stereo Rasterization}
\label{sec:stereo}

\Sect{sec:ch:op} shows the strong similarity between left- and right-eye images.
This section describes our stereo rendering pipeline, which leverages the stereo similarity between the two eyes to reduce the rendering computations.

\para{Motivation.}
While prior studies~\cite{feng2024cicero, chaurasia2020passthrough+, vona2025comparing} have proposed methods to exploit stereo similarity between two eye images, these techniques have two key limitations.
First, existing methods rely on a high-fidelity depth map to perform accurate warping; however, the depth maps produced by 3DGS are often unreliable.
Second, directly warping pixels from the left eye to the right eye compromises the view-dependent characteristic of 3DGS.
It often produces less photo-realistic images, as mentioned in prior work~\cite{feng2024cicero}.

To address these issues, we propose a triangulation-based technique that skips redundant computation in the remaining stages, i.e., preprocessing, sorting, and rasterization, while preserving \textit{bit-accurate} results in 3DGS.

\begin{figure}[t]
    \centering
    \includegraphics[width=0.6\columnwidth]{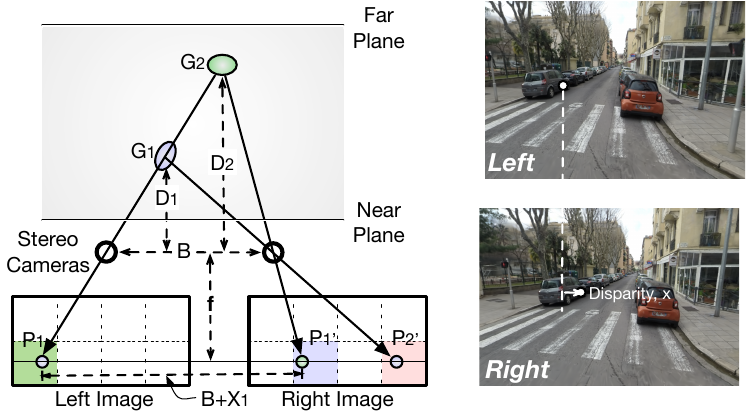}
    \caption{The intuition behind our stereo rasterization, which leverages the triangulation technique~\cite{hartley2003multiple, szeliski2010computer}. 
    For each Gaussian, once we determine the pixels it intersects in the left-eye image, we can directly compute, via triangulation, the corresponding pixel locations it will contribute to in the right-eye image, without preprocessing and sorting. }
    \label{fig:stereo_vision}
\end{figure}

\para{Intuition.}
We first give a toy example in \Fig{fig:stereo_vision} to show the intuition behind our algorithm: \textit{given the Gaussian relative depth to the left camera and the geometry of the stereo camera, we can directly compute which pixel in the right image would be contributed by this Gaussian point.}

\Fig{fig:stereo_vision} shows a stereo camera with a baseline of $B$, defined as the horizontal distance between the left and right cameras in a VR headset.
Both cameras have the same focal length, $f$.
Consider a pixel $P_1$ in the left-eye image, which is contributed to by two Gaussians, $G_1$ and $G_2$.
The depths of $G_1$ and $G_2$ to the camera center are $D_1$ and $D_2$, respectively.

Based on \textit{triangulation}~\cite{hartley2003multiple, szeliski2010computer}, a widely-known process in computer vision, we calculate the \textit{disparity}, $X_1 = P_1' - P_1$, i.e., the horizontal displacement between the two corresponding pixels $P_1'$ and $P_1$ in the left and right images as, $X_1 = Bf/D_1.$
Similarly, the disparity for $G_2$ is given by $X_2 = Bf/D_2$.
With these disparity results, we can determine, in the right image, which pixels each Gaussian will contribute to, without re-running the preprocessing and sorting stages.

In 3DGS rendering, a near plane and a far plane are defined to avoid rendering artifacts~\cite{kerbl20233d}.
Given the near-plane distance, the maximum disparity in a typical VR setup is bounded within 16 pixels, since disparity is inversely proportional to depth.

\para{Algorithm.}
In \Sect{sec:bg}, we explain that 3DGS pipelines render in a tile-by-tile fashion, specifically, $4\times4$ granularity.
Our algorithm also adapts the tile-based rendering and maps every Gaussian contributing to a tile in the left-eye image to its corresponding tile in the right-eye image.
For example, $G_1$, which contributes to the green tile in the left image, is mapped to the pink tile in the right-eye image in \Fig{fig:stereo_vision}. 
We now describe the changes in the rendering pipeline.

\begin{figure*}[t]
    \centering
    \includegraphics[width=\textwidth]{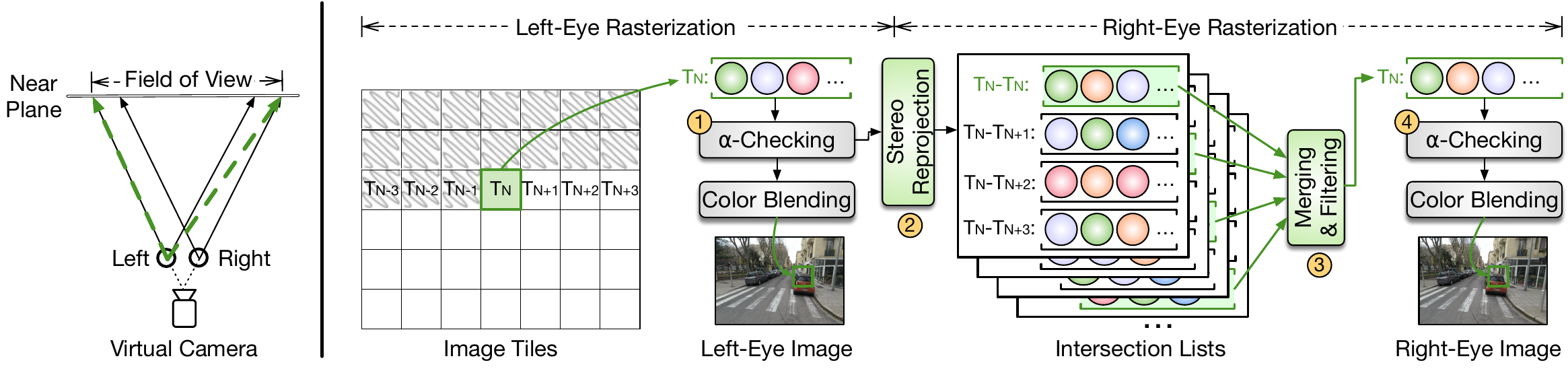}
    \caption{Overview of stereo rasterization. Left: for preprocessing and sorting, we use a wider FoV to cover the FoVs of both eyes. Right: for rasterization, we first perform a standard rasterization to render the left-eye image; for the right-eye image, instead of reprocessing all Gaussians, we leverage the geometric relationship between Gaussians and the stereo camera, and map those contributing Gaussians to the right view via triangulation. Thus, we largely reduce the redundant computations.
}
    \label{fig:stereo_rasterization}
\end{figure*}

\subparagraph{Preprocessing\&Sorting.}
The left part of \Fig{fig:stereo_rasterization} shows our key modification to the preprocessing and sorting stages.
That is to allow the left- and right-eye images to share these computations of these two stages, given their highly similar fields of view (FoVs).
To do that, we place a virtual camera slightly behind both eyes to determine the common FoV between them.
Then, we derive the left-eye FoV that fully covers the common region (the green bashed FoV), and perform preprocessing and sorting on this FoV to avoid repetitively computing the preprocessing and sorting twice.

\subparagraph{Stereo Rasterization.}
The right part of \Fig{fig:stereo_rasterization} shows our key contribution, \textit{bit-accurate} stereo rasterization pipeline:

\circnum{1} 
We first render the left‑eye image following the standard rasterization process.
For a given tile $T_N$, each pixel iterates over all Gaussians in the $T_N$ list, which is a sorted list after sorting (see \Fig{fig:pipeline}). 
For each Gaussian, each pixel first performs an $\alpha$-check: if the transmittance exceeds the threshold $\alpha^{*}$, the Gaussian’s color is blended into the pixel; otherwise, the Gaussian is skipped.
The final pixel value is obtained after processing all Gaussians in the list.

\circnum{2}
If a Gaussian passes an $\alpha$-check, it definitely contributes to the right-eye image.
We then apply triangulation, as described in \Fig{fig:stereo_vision}, to transform this Gaussian to the right-eye view.
Based on the computed disparity, we can determine which tile in the right-eye image this Gaussian intersects, and insert this Gaussian into the corresponding list.
Based on the maximum disparity (16 pixels), this Gaussian will be inserted into one of four lists, i.e., $T_N$-$T_N$, $T_N$-$T_{N+1}$, $T_N$-$T_{N+2}$, and $T_N$-$T_{N+3}$.
$T_N$-$T_{N+1}$ stands for the Gaussians from tile $T_N$ in the left image that would contribute to tile $T_{N+1}$ in the right image.
Each tile maintains its four intersection lists.

\circnum{3}
To render tile $T_N$ in the right-eye image, we first need to identify which Gaussians intersect with this tile.
Based on the triangulation, the intersected Gaussians can only come from four lists: $T_{N-3}$-$T_N$, $T_{N-2}$-$T_N$, $T_{N-1}$-$T_N$, and $T_N$-$T_N$.
The complete intersection set for $T_N$ is obtained by merging these four lists.
Since each list is already sorted, we can skip re-sorting these four lists.
Instead, we directly merge them and remove duplicate ones.
This process is analogous to the merge phase of merge sort, but with four pre-sorted lists.

\circnum{4}
Once we obtain the complete and sorted intersection list for $T_N$, we can render tile $T_N$ of the right-eye image following the same procedure as for the left-eye image.
Since our pipeline integrates the same set of contributing Gaussians as the original pipeline, the rendering result is bit-accurate.

In this process, the Gaussians processed by the right-eye image have already passed the $\alpha$-check; thus, our stereo rasterization inherently avoids part of rasterization workload for right eyes and achieves 1.4$\times$ speedup on mobile GPUs.

\section{Architectural Support}
\label{sec:arch}

\begin{figure}[t]
    \centering
    \includegraphics[width=\columnwidth]{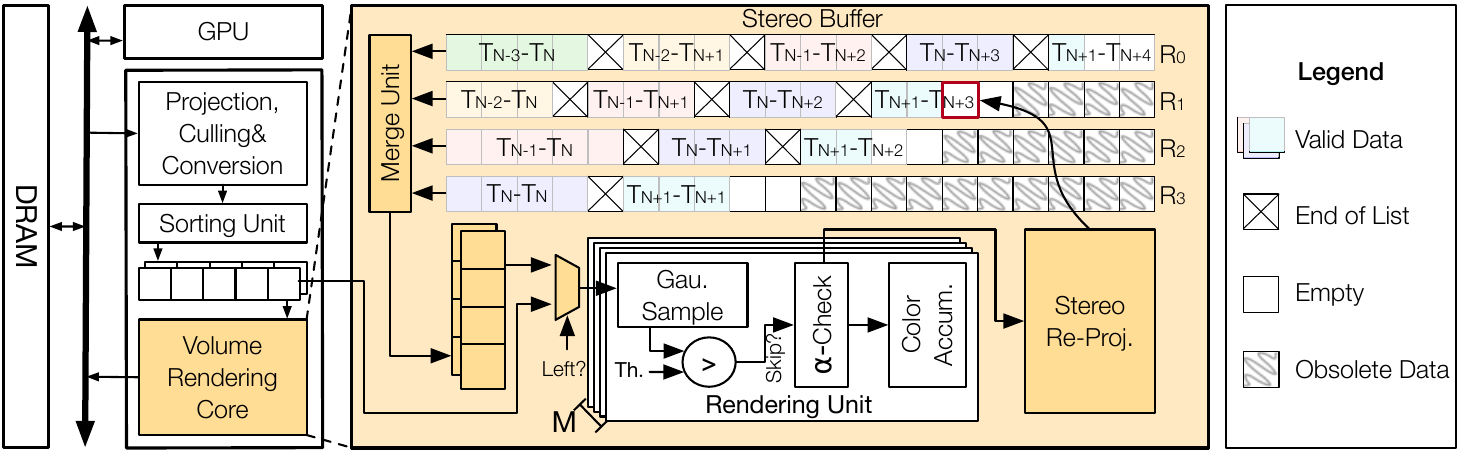}
    \caption{The overview architecture design. We augmented the basic architecture, GSCore~\cite{lee2024gscore}, to support stereo rasterization (colored in yellow).}
    \label{fig:arch}
\end{figure}

Supporting stereo rasterization in \Sect{sec:stereo} on existing 3DGS accelerators requires only minimal hardware augmentation.
In this section, we use GSCore~\cite{lee2024gscore} as an example to illustrate the necessary modifications.
Other architectures can adapt our technique in a similar fashion.

\para{Overview.}
\Fig{fig:arch} illustrates the overall pipelined architecture of \proj, which integrates a local GPU on the SoC with a GSCore-like accelerator~\cite{lee2024gscore}.
In our design, the GPU executes LoD search and identifies the visible Gaussians at the appropriate level of detail.
The runtime Gaussian management system then fetches the corresponding rendering attributes and streams them to the accelerator for the remaining rendering stages.
The accelerator follows the GSCore pipeline, where the projection unit, sorting unit, and volume rendering core (VRC) are pipelined to render different tiles within a frame.
We augment the VRC to support stereo rasterization, allowing the left- and right-eye views in VR to share computation and reduce redundant memory accesses.

\para{Support for Stereo Rasterization.}
\Fig{fig:arch} highlights our modified components in colors.
The basic VRC consists of $M$ rendering units (RUs).
Each RU is responsible for rendering one pixel.
For each Gaussian, its attributes are first broadcast to all RUs.
Based on the result of the $\alpha$-check, each RU determines whether the Gaussian contributes to its pixel and, if so, blends it into the accumulated color.

Meanwhile, all the results of the $\alpha$-check are forwarded to our augmented stereo re-projection unit (SRU).
If any pixel within the tile integrates this Gaussian, then the SRU would re-project this Gaussian into the right-eye view.
Based on the re-projected disparity, the SRU would write the Gaussian into the corresponding list in the stereo buffer.


Our stereo buffer adopts the classic line buffer design from image processing~\cite{whatmough2019fixynn, chi2018soda, hegarty2014darkroom, ujjainkar2023imagen}, as shown in \Fig{fig:arch}.
To avoid the bank conflicts, each row stores a single disparity category.
For instance, row $R_0$ stores Gaussians whose disparity is greater than 3 tiles (i.e., 12 pixels) between the left and right eyes.
Here, $T_{N-3}$-$T_{N}$ denotes the Gaussians from tile $T_{N-3}$ in the left image that would contribute to tile $T_{N}$ in the right image.
Each time, SRU writes a Gaussian to one of the rows based on the disparity result.
Meanwhile, the merge unit sorts the current four lists for $T_N$ in the right image, i.e., $T_{N-3}$-$T_{N}$, $T_{N-2}$-$T_{N}$, $T_{N-1}$-$T_{N}$, and $T_{N}$-$T_{N}$, by reading the head entries of the four rows and selecting the minimum.
Each row is designed as a circular buffer to maximize utilization.

\para{Pipelining.}
Similar to GSCore~\cite{lee2024gscore}, our architectural design pipelines the three stages, preprocessing, sorting, and rasterization, and renders image tiles in row-major order.
In our stereo rasterization, we sequentially render the corresponding tiles in the left-eye and right-eye images.
Note that, right-eye tiles begin to render after left-eye tile rendering, starting from the fourth tile.
The first three tiles in the right-eye images are rendered independently.

\section{Experimental Setup}
\label{sec:exp}

\para{Hardware Implementation.}
We develop a RTL implementation of \proj clocked at 1 GHz, where the basic configuration is similar to GSCore~\cite{lee2024gscore}.
\proj consists of four projection units, four hierarchical sorting units, and eight VRCs.
Each VRC consists of $4\times4$ rendering units and a 16~KB feature buffer.
We augment each VRC with one stereo reprojection unit, one merge unit, and a 16~KB stereo buffer banked at 4~KB granularity.
In addition, a 144~KB global double buffer is used to store the intermediate data of the pipeline.
Our RTL design is implemented via Synopsys synthesis and Cadence layout tools in TSMC 16nm FinFET technology.
SRAMs generated by an ARM memory compiler. 
Power is simulated using Synopsys PrimeTimePX, with fully annotated switching activity.
The DRAM is modeled after 4 channels of Micron 16~Gb LPDDR3-1600 memory~\cite{micronlpddr3}.
DRAM energy is calculated using Micron's System Power
Calculators based on the memory traffic~\cite{microdrampower}.
The numbers of our RTL designs are then scaled down to 8 nm node using DeepScaleTool~\cite{stillmaker2017scaling, sarangi2021deepscaletool} to match the mobile Ampere GPU on Nvidia Orin~\cite{orinsoc}.

\para{Simulation Methodology.}
We simulate the entire system with our cycle-accurate simulator, which is implemented with component-level latencies and power measurements.
The latency and energy of VRC are obtained from the post-synthesis results of its RTL design and scaled down to 8 nm node using DeepScaleTool~\cite{stillmaker2017scaling, sarangi2021deepscaletool} to match the mobile Ampere GPU on Nvidia Orin~\cite{orinsoc}.
The latency of GPU execution is measured, including kernel launch time.
The GPU power consumption is obtained via the built-in power measurement on the Nvidia Orin SoC.
The system energy is the sum of GPU, VRC, and DRAM. 
We then build a cycle-level simulator of the architecture with the latency of each component parameterized from measurements (for GPU) and post-synthesis results of the NPU design.

\para{Area.}
\proj introduces a minimal area overhead compared to the baseline architecture.
The main overhead comes from additional 16~KB SRAM required for each VRC.
Overall, the additional hardware introduces around 14\% area overhead (0.25~mm$^2$), compared to GSCore (1.78~mm$^2$) in 16nm.

\para{Software Setup.}
We evaluate on three large-scale datasets: Urban~\cite{lin2022capturing}, Mega~\cite{turki2022mega}, and HierGS~\cite{kerbl2024hierarchical}, as well as three small-scale datasets: T\&T~\cite{Knapitsch2017}, DB~\cite{hedman2018deep}, and M360~\cite{barron2022mipnerf360}.
To assess the effectiveness of \proj, we evaluate against three large-scale 3DGS algorithms: HierGS~\cite{kerbl2024hierarchical}, CityGS~\cite{liu2024citygaussian}, and OctreeGS~\cite{ren2024octree}.
The main difference of those algorithms is the LoD search.
For rendering quality, we adopt two widely used metrics: peak signal-to-noise ratio (PSNR) and structural similarity index (SSIM).
To mimic real VR scenarios, all stereo images are rendered at $2064\times2208$ resolution with a pupil baseline of 6~cm.

\para{Software Baselines.}
To evaluate the quality of our stereo rasterization, we compare against three baselines:
\begin{itemize}
    \item \mode{Base}: the baseline algorithm that renders both eyes.
    \item \mode{Warp}~\cite{chaurasia2020passthrough+}: a widely-used warping technique which uses a classic densification to fill disocclusions.
    \item \mode{Cicero}~\cite{feng2024cicero}: a state-of-the-art warping-based method, which uses neural rendering to fill disocclusions.
\end{itemize}
Note that, both \mode{Warp} and \mode{Cicero} use the generated depth map from 3DGS~\cite{chung2024depth} rather than the ground truth depth, which is not available in real-world scenarios.

\para{Hardware Baselines.}
To evaluate the efficiency of our hardware design, we compare three hardware baselines:
\begin{itemize}
    \item \mode{GPU}: a mobile Ampere GPU on Nvidia Orin SoC~\cite{orinsoc}.
    \item \mode{GSCore}~\cite{lee2024gscore}: a dedicated accelerator for 3DGS.
    \item \mode{GBU}~\cite{ye2025gaussian}: a hardware module that accelerates rasterization while the remaining operations are executed on the same mobile GPU as \mode{GPU}. For fairness, we implement 128 Row PEs in GBU to align with GSCore. 
    
\end{itemize}
We develop the RTL implementation of both \mode{GBU} and \mode{GSCore}. 
The numbers of both RTL designs are then scaled down to 8~nm node to match Nvidia Orin~\cite{orinsoc}.

\section{Evaluation}
\label{sec:eval}

\subsection{Rendering Quality}
\label{sec:eval:acc}

\begin{figure}[t]
\centering
\subfloat[PSNR evaluation.]{
	\label{fig:stereo_psnr}
    \includegraphics[width=0.49\columnwidth]{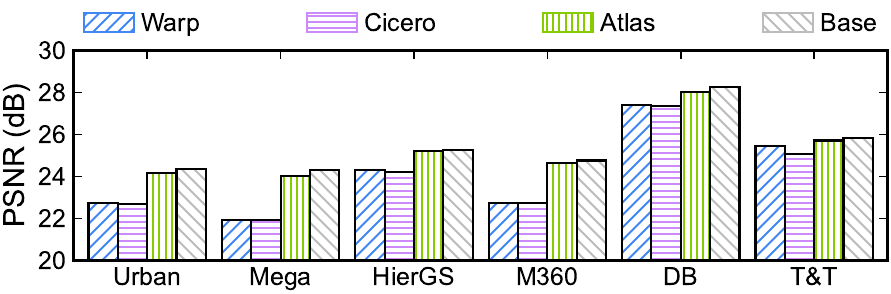}
    }
\subfloat[SSIM evaluation.]{
	\label{fig:stereo_ssim}
	\includegraphics[width=0.49\columnwidth]{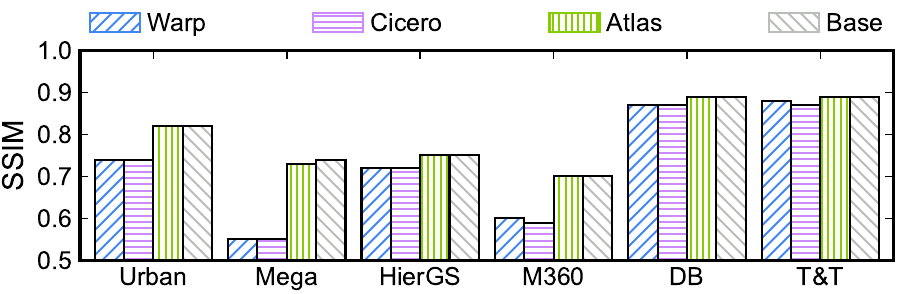}
}
\caption{Rendering quality evaluation on stereo warping.}
\label{fig:stereo_warp}
\end{figure}

\para{Stereo Rendering.}
We first evaluate the end-to-end visual quality of stereo rendering, as shown in \Fig{fig:stereo_warp}.
Here, \mode{Base} renders both eyes using HierGS~\cite{kerbl2024hierarchical}, while both \mode{Warp} and \mode{Cicero} generate the right-eye view by warping the left-eye image.
As expected, both \mode{Warp} and \mode{Cicero} introduce noticeable accuracy loss against \mode{Base}.
In contrast, \mode{\proj} delivers nearly identical quality to \mode{Base}, with only a 0.1~dB PSNR loss.
In terms of SSIM, \mode{\proj} shows no quality loss.
Meanwhile, this minor accuracy loss is not introduced by our stereo rasterization, which is bit-accurate, but by the on-demand loading of our runtime Gaussian management and the LoD search performed once every $w$ frames, which may temporarily leave a few Gaussians unavailable.

\subsection{GPU Performance}
\label{sec:eval:gpu}
We first give the GPU performance comparison by rendering two eyes at $2064\times2208$ pixels per eye in VR.
The performance comparison with our dedicated hardware support is shown in \Sect{sec:eval:perf}.

\begin{figure}[t]
\centering
\begin{minipage}[t]{0.49\columnwidth}
    \centering
    \includegraphics[width=1\columnwidth]{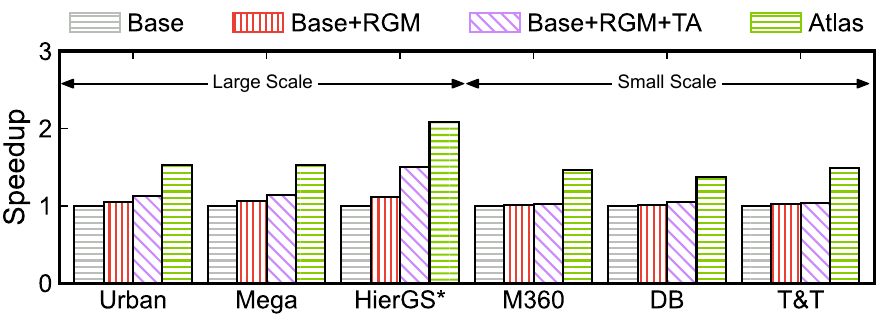}
    \caption{GPU speedup of the algorithmic optimizations in \proj. All numbers are normalized to \mode{Base}.}
    \label{fig:gpu_speedup}
\end{minipage}
\hspace{2pt}
\begin{minipage}[t]{0.49\columnwidth}
  \centering
  \includegraphics[width=\columnwidth]{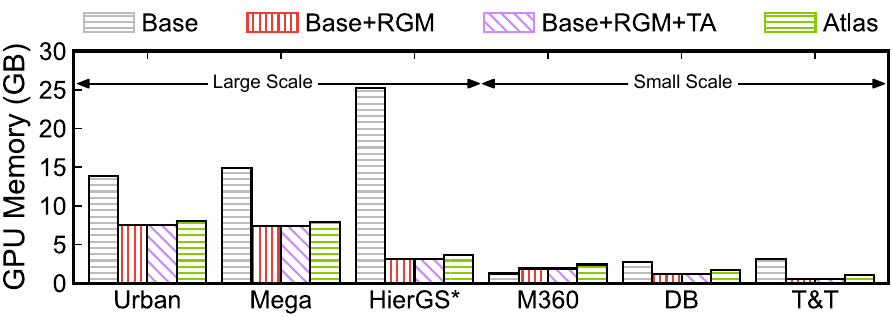}
  \caption{Overall GPU memory footprint. \proj reduces memory usage across all large-scale scenes.}
  \label{fig:gpu_mem}
\end{minipage}
\end{figure}

\para{GPU Speedup.}
\Fig{fig:gpu_speedup} shows the GPU-side speedup of the algorithmic optimizations in \proj.\footnote{The asterisk on HierGS indicates that some HierGS scenes are too large to be rendered by \mode{Base} on the mobile GPU and are therefore excluded from the reported measurements.}
We evaluate four variants: \mode{Base}, which executes the HierGS pipeline; \mode{Base+RGM}, which additionally applies the runtime Gaussian management (\Sect{sec:mem}); \mode{Base+RGM+TA}, which further applies the temporal-aware LoD search (\Sect{sec:lod}); and \mode{\proj}, which applies all optimizations.
All numbers are normalized to the \mode{Base} latency. Overall, \mode{\proj} achieves the highest GPU-side speedup, up to 2.1$\times$, compared to \mode{Base}.

\para{Memory Footprint.}
\Fig{fig:gpu_mem} compares the GPU memory footprint across the four variants of \proj. With the runtime Gaussian management mechanism, \proj reduces the footprint by up to 7.0$\times$ compared to \mode{Base} on large-scale scenes.
This keeps the memory demand of all large-scale scenes below 12~GB, the capacity of mainstream VR devices.
The reduction is most pronounced on the HierGS~\cite{kerbl2024hierarchical} dataset, whose street-level viewpoints activate far fewer spatial blocks than the aerial viewpoints of Urban~\cite{lin2022capturing} and Mega~\cite{turki2022mega}.
Note that, the GPU implementation of the stereo rasterization requires additional GPU memory, but the extra footprint is modest.

\subsection{Accelerator Performance and Energy}
\label{sec:eval:perf}

\begin{figure}[t]
  \centering
  \begin{minipage}[t]{0.49\columnwidth}
    \centering
    \includegraphics[width=\columnwidth]{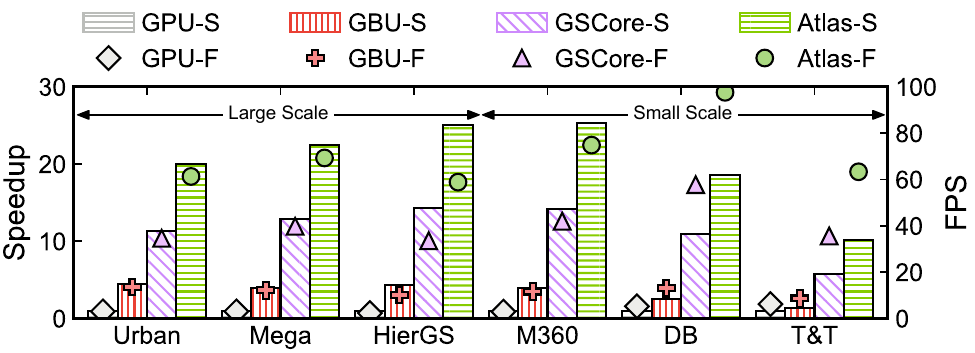}
    \caption{Hardware performance comparison. All numbers are normalized to \mode{GPU}. ``S'': speedup; ``F'': frame per second.}
    \label{fig:overall_speedup}
  \end{minipage}
  \hspace{2pt}
  \begin{minipage}[t]{0.49\columnwidth}
    \centering
    \includegraphics[width=\columnwidth]{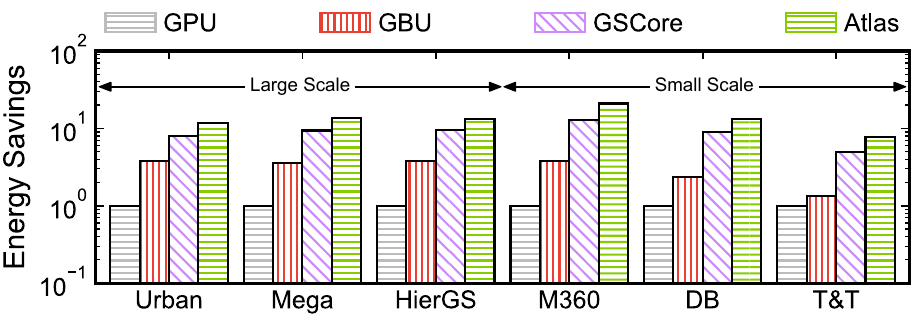}
    \caption{Hardware overall energy. Display power is excluded since it is constant across methods. Numbers are normalized to \mode{GPU}.}
    \label{fig:overall_energy}
  \end{minipage}
\end{figure}

\begin{figure}[t]
\centering
\begin{minipage}[t]{0.32\columnwidth}
  \centering
  \includegraphics[width=\columnwidth]{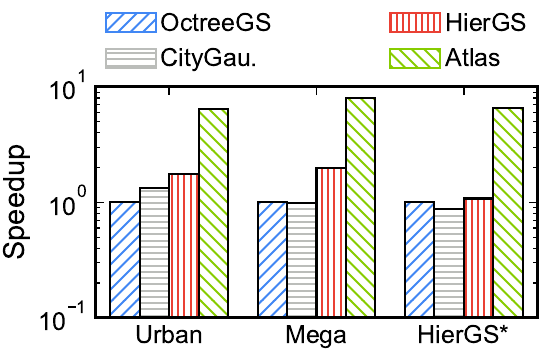}
  \caption{Speedup on LoD search. Temporal-aware LoD search achieves better performance than prior methods~\cite{ren2024octree, kerbl2024hierarchical, liu2024citygaussian}.}
  \label{fig:lod_speedup}
\end{minipage}
\hspace{2pt}
\begin{minipage}[t]{0.32\columnwidth}
  \centering
  \includegraphics[width=\columnwidth]{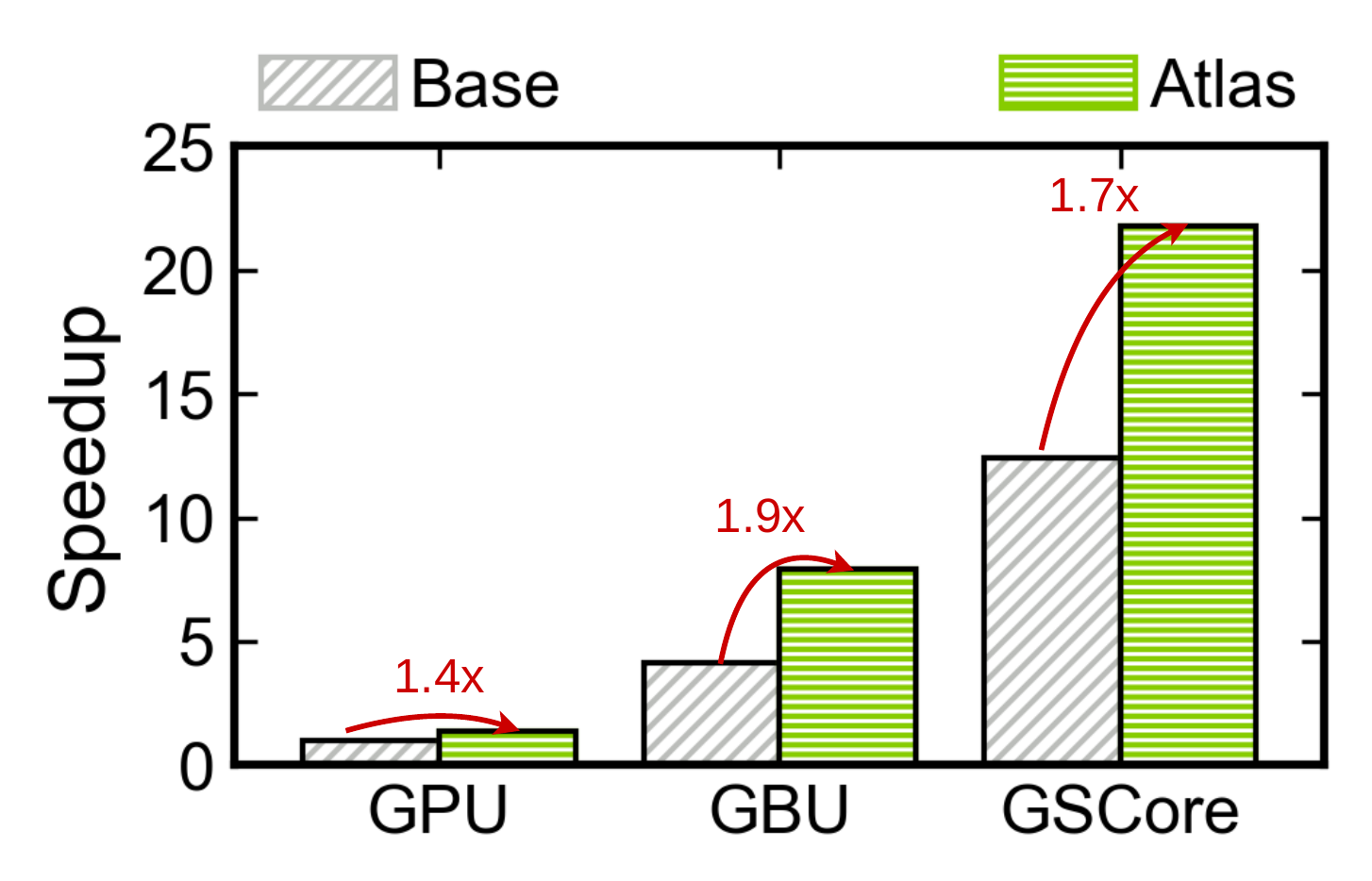}
  \caption{The speedup of \proj against the baseline algorithm on the local device. Performance numbers are normalized to \mode{GPU}.}
  \label{fig:raster_speedup}
\end{minipage}
\hspace{2pt}
\begin{minipage}[t]{0.32\columnwidth}
  \centering
  \includegraphics[width=\columnwidth]{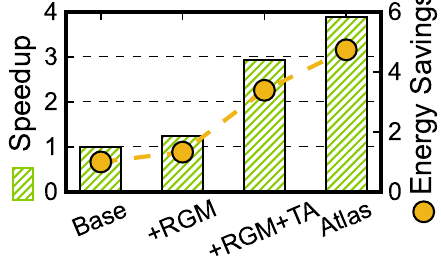}
  \caption{Ablation study on \proj. \mode{RGM}: apply runtime Gaussian management; \mode{TA}: apply temporal-aware LoD search; \mode{SR}: apply stereo rasterization. \proj applies all three.}
  \label{fig:ablation}
\end{minipage}
\end{figure}

\para{Speedup.}
\Fig{fig:overall_speedup} presents the overall performance comparison.
All variants execute HierGS~\cite{kerbl2024hierarchical} since it has the best LoD search performance in \Fig{fig:lod_speedup}.
Except for \mode{\proj}, which uses the stereo rasterization in \Sect{sec:stereo}, all other methods render the left and right eyes in two separate passes.
We report the motion-to-photon latency, normalized to the \mode{GPU} baseline.

On average, \mode{\proj} achieves the highest speedup, 18.5$\times$, compared to \mode{GPU}.
Meanwhile, we also show the frame rate of different methods.
Here, we assume rendering and data communication can be pipelined, similar to prior work~\cite{xie2021q}.
Overall, \mode{\proj} achieves 70.9~FPS.
While \mode{\proj} does not achieve the VR requirement, 90~FPS, \Sect{sec:eval:sens} shows that \proj can easily achieve 90~FPS by scaling up VRC.

\para{Energy Savings.}
\Fig{fig:overall_energy} shows the overall energy savings against \mode{GPU}.
Among all methods, \mode{\proj} delivers the best efficiency, achieving 1.5$\times$ and 13.1$\times$ lower energy on average compared to \mode{GSCore} and \mode{GPU}, respectively.

\para{LoD Search.}
\Fig{fig:lod_speedup} shows the performance comparison of different algorithms on LoD search. As in \Sect{sec:eval:gpu}, the asterisk on HierGS indicates that scenes too large for the mobile GPU baseline are excluded.
The original LoD search in OctreeGS~\cite{ren2024octree} is used as the baseline, and we compare CityGau~\cite{liu2024citygaussian} and HierGS~\cite{kerbl2024hierarchical}.
\mode{\proj} achieves much higher speedup (up to 8.1$\times$) than other methods by exploiting temporal similarity to eliminate redundant node accesses.

\para{Local Rendering.}
\Fig{fig:raster_speedup} shows the speedup of our stereo rasterization on local rendering (including preprocessing, sorting, and rasterization) over six datasets.
Across all architectural designs, \proj consistently delivers 1.4$\times$, 1.9$\times$, and 1.7$\times$ speedups on \mode{GPU}, \mode{GBU}, and \mode{GSCore}, respectively.

\subsection{Ablation Study}
\label{sec:eval:abl}

\Fig{fig:ablation} shows the ablation study of the contributions in \proj.
\mode{Base} executes the HierGS algorithm on our architecture.
We show the speedup (left y-axis) and energy savings (right y-axis) under:
1) \mode{Base} with only our runtime Gaussian management (RGM),
2) \mode{Base} with RGM and temporal-aware LoD search (TA),
3) \mode{Base} with all optimizations.
\mode{RGM} alone achieves 1.3$\times$ speedup and 1.3$\times$ energy savings, as on-demand loading confines LoD search to the active blocks instead of the entire scene, reducing the LoD search overhead.
Adding \mode{TA} achieves 2.9$\times$ speedup and 3.4$\times$ energy savings since LoD search dominates the end-to-end latency of large-scale scenes and \mode{TA} eliminates most redundant tree accesses by exploiting temporal similarity.
All together, \mode{\proj} achieves 3.9$\times$ speedup and 4.7$\times$ energy savings.

\subsection{Sensitivity Study}
\label{sec:eval:sens}

\begin{figure}[t]
\centering
\begin{minipage}[t]{0.32\columnwidth}
  \centering
  \includegraphics[width=\columnwidth]{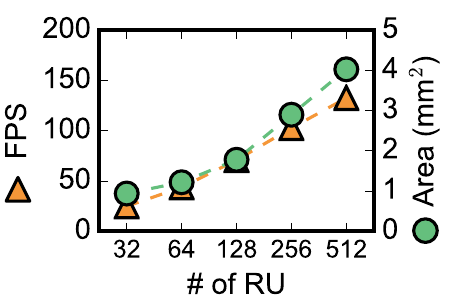}
  \caption{The scalability of performance and area to the number of rendering unit in VRC. \proj can easily achieve 90~FPS by scaling up.}
  \label{fig:ru_scalability}
\end{minipage}
\hspace{2pt}
\begin{minipage}[t]{0.32\columnwidth}
  \centering
  \includegraphics[width=\columnwidth]{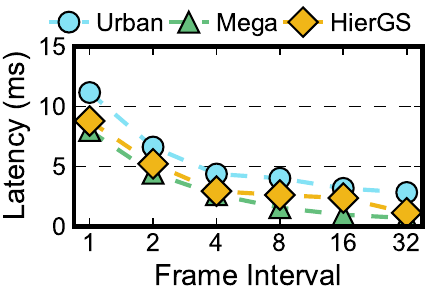}
  \caption{The sensitivity of the average per-frame data-transfer latency to the frame interval, $w$, under three large-scale datasets.}
  \label{fig:frame_interval}
\end{minipage}
\hspace{2pt}
\begin{minipage}[t]{0.32\columnwidth}
  \centering
  \includegraphics[width=\columnwidth]{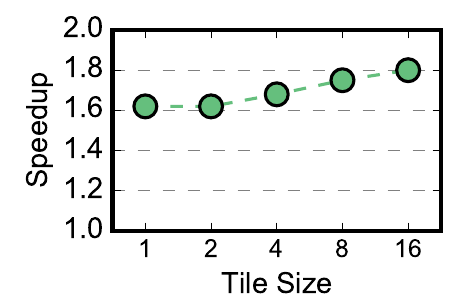}
  \caption{The sensitivity of performance to the tile size.}
  \label{fig:tile_size}
\end{minipage}
\end{figure}

\para{RU Scalability.}
We demonstrate that \proj can readily meet VR frame rate requirements by scaling the rendering units (RUs) in the VRC. 
\Fig{fig:ru_scalability} shows the performance trend by averaging the results from three large-scale datasets~\cite{kerbl2024hierarchical, lin2022capturing, turki2022mega}. 
We show that doubling the RUs from 128 to 256 in the default VRC configuration enables our architecture to achieve real-time VR performance. 
However, increasing from 128 to 256 RUs increases the area by 62.9\%.

\para{Frame Interval, $w$.}
\Fig{fig:frame_interval} shows the sensitivity of the average per-frame data-transfer latency to $w$ under three large-scale datasets.
As $w$ increases, the average per-frame data-transfer latency decreases monotonically, since the on-demand loading of Gaussian attributes is amortized over more frames.
Moving from $w = 1$ to $w = 4$ already reduces the latency by 2.5--3.0$\times$, confirming that most Gaussians selected by LoD search are reused across adjacent frames.
We choose $w = 4$ as the default as it is enough to hide the rendering and data-transfer latency, yet remains the visual inconsistency between consecutive frames.

\para{Tile Size.}
Lastly, \Fig{fig:tile_size} shows the sensitivity of rendering performance to the tile size using HierGS dataset~\cite{kerbl2024hierarchical}.
The speedup is normalized to the corresponding baseline with the same tile size.
We observe that the speedup decreases modestly as the tile size decreases. With smaller tiles, a Gaussian's projected footprint spans more tiles, so the same Gaussian is inserted into and merged from more per-tile lists in stereo rasterization, which adds redundant overhead.

\section{Related Work}
\label{sec:related}

\para{3DGS Acceleration.}
Recent studies propose various 3DGS architectures~\cite{feng2025lumina, ye2025gaussian, lee2024gscore, li2025uni, lee2025vr, lin2025metasapiens, durvasula2025arc, he2025gsarch, huang2026splatonic, li2023sltarch, zhang2025streaming, zhu2026nebula, xu2025animatable, liu2025comprehensive, pei2026degs, li2026orange}.
A few studies~\cite{lee2024gscore, ye2025gaussian, lin2025metasapiens, feng2025lumina, lee2025vr} are designed for the acceleration of the forward pass in 3DGS.
For instance, MetaSapiens~\cite{lin2025metasapiens} and GBU~\cite{ye2025gaussian} address the workload imbalance during rasterization. 
Lumina~\cite{feng2025lumina} proposes a caching technique to avoid redundant computation.
Some propose solutions for 3DGS training~\cite{he2025gsarch, durvasula2025arc}.
For example, ARC~\cite{durvasula2025arc} addresses the atomic operations in training while GSArch~\cite{he2025gsarch} prunes redundant gradient updates.
In principle, \proj is orthogonal to those works and can be applied to any existing 3DGS acceleration framework to support large-scale 3DGS rendering.

\para{Collaborative Rendering.}
There are a few cloud-client collaborative rendering methods\mbox{~\cite{xie2021q, xu2023edge, he2020collabovr, ke2023collabvr, leng2019energy, feng2024cicero, zhao2020deja, chen2025baft}} in literature. 
However, all current techniques are designed for mesh-based rasterization pipelines, not for 3DGS. 
For instance, Cicero\mbox{~\cite{feng2024cicero}} and CollabVR\mbox{~\cite{ke2023collabvr}} offload all rendering tasks to the cloud and only perform lightweight warping at the client side to accommodate disocclusions or stereo display. 
Both E-VR\mbox{~\cite{leng2019energy}} and DejaView\mbox{~\cite{zhao2020deja}} focus on 360 video streaming. 
Both leverage the unique features in 360 videos, e.g., the spatio-temporal redundancies or the user's field of view, to reduce the reprojection overhead and network traffic. 
Meanwhile, both Q-VR\mbox{~\cite{xie2021q}} and EDC\mbox{~\cite{xu2023edge}} incorporate foveated rendering to reduce the on-device workload.

However, all these methods continue to face bandwidth limitations when targeting higher FPS/resolution. 
More importantly, they partition the workload at the pixel level, which is incompatible with 3DGS since clients still require compute-intensive LoD search to filter Gaussians.
We propose \proj, a novel cloud-client co-design framework tailored for 3DGS and substantially reduce the on-device workload while minimizing the network traffic.

\para{Warping Techniques.}
Image warping~\cite{chaurasia2020passthrough+, szeliski2022image} is a lightweight technique for synthesizing novel views in conventional image-based rendering~\cite{chen2023view, chen1995quicktime}.
It exploits spatial and temporal correlations across frames to avoid redundant computations~\cite{buckler2018eva2, feng2020real, zhu2018euphrates, feng2023fast, ying2022exploiting, feng2019asv, zhao2020deja, zhao2021holoar, zhang2026dstar, miao2026kaleido, feng2022real}.
\proj also leverages both those similarities, but unlike prior work, \proj achieves bit-accurate rendering while delivering speedup.

\section{Conclusion}
\label{sec:conc}

Human imagination is boundless, and so too should be virtual 3D Gaussian worlds.
\proj marks the first step toward real-time, infinite-scale 3DGS splatting in VR.
By offloading the majority of Gaussians to the disk storage, we introduce a rendering framework that alleviates communication bottlenecks and a novel stereo rasterization pipeline that eliminates redundancy in VR stereo rendering.


\bibliographystyle{ACM-Reference-Format}
\bibliography{reference}

\end{document}